\documentclass[11pt,a4paper]{article}
\usepackage{jhepus}

\usepackage{amsfonts,amsbsy,latexsym,amssymb,amscd,amstext}
\usepackage{epsfig}
\usepackage{graphicx}
\newcommand{\be}{\begin{equation}}
\newcommand{\ee}{\end{equation}}
\newcommand{\ba}{\begin{array}}
\newcommand{\ea}{\end{array}}
\newcommand{\baa}{\begin{array}}
\newcommand{\eaa}{\end{array}}
\newcommand{\bea}{\begin{eqnarray}}
\newcommand{\eea}{\end{eqnarray}}
\newcommand{\half}{\frac{1}{2}}

\newcommand{\sig}{\sigma}
\newcommand{\bsig}{\overline{\sigma}}

\newcommand{\II}{\mathbf{I}}

\newcommand{\Op}{O^{(+)}}
\newcommand{\Opl}{O^{(+)}_L}
\newcommand{\Opm}{O^{(\pm)}}

\newcommand{\Ovia}{O_{IA}}
\newcommand{\AFM}{\rm afm}

\title{Ultraviolet filtering of lattice configurations and
applications to Monte Carlo dynamics }

\author[a]{Margarita Garc\'{\i}a P\'erez,}
\author[a,b]{Antonio Gonz\'alez-Arroyo}
\author[a]{and Alfonso Sastre}

\affiliation[a]{Instituto de F\'{\i}sica Te\'orica UAM/CSIC}
\affiliation[b]{Departamento de F\'{\i}sica Te\'orica \\
       Universidad Aut\'onoma de Madrid, E-28049--Madrid, Spain}

\emailAdd{margarita.garcia@uam.es}
\emailAdd{antonio.gonzalez-arroyo@uam.es}
\emailAdd{alfonso.sastre@uam.es}

\abstract{
We present a detailed study of a filtering method based upon Dirac
quasi-zero-modes in the adjoint representation. The procedure induces
no distortions on configurations which are 
solutions of the euclidean classical equations of motion. On the other hand, it
is very effective in reducing the short-wavelength stochastic noise
present in Monte-Carlo generated configurations. After testing the
performance of the method in various situations, we apply it
successfully to study the effect of Monte-Carlo dynamics on
topological structures like instantons.
}


\begin{document}
\maketitle

\section{Introduction}
\label{s.intro}
In Quantum Field Theory the path integral is supported on distributions.
Nevertheless, the  semiclassical  method aims at  obtaining a
reduced path integral space,  labelled by smooth configurations. This
is not in contradiction since the weights account for all the stochastic 
fluctuations around them. In the weak  coupling limit this methodology works
successfully for many theories. In the case of QCD,  one might
consider, for example,  a dilute gas of instantons parametrised by
its positions and sizes~\cite{Callan:1977gz}. Classically, the distribution is Poissonian with a uniform distribution
in space, a scale invariant distribution in sizes and a classical weight for each 
new instanton which is the exponential of minus its action. This is not
the  appropriate path integral measure though, since the integration over
gaussian  fluctuations around an instanton forces to replace the action by the
free energy. This induces a size dependence dictated by the renormalization group.
Nevertheless, such a reduced path-integral space, although successful
in predicting a non-zero topological susceptibility, crucial for the
solution of the U(1) problem, does not reproduce other phenomena
believed to be present in the full theory, such as Confinement.
Presumably a picture based upon the classical vacuum ($A_\mu=0$) with
isolated lumps is a poor description of the true vacuum. Indeed, the quantum weights
determined by perturbation theory already tell us that instantons
prefer to grow and overlap with others~\cite{'tHooft:1976fv,Belavin:1976vj}. 
A possible modified scenario, in which instantons are close to each other but
preserve their individuality is the {\em instanton liquid model}~\cite{Shuryak:1981ff}-\cite{Schafer:1996wv}.
In any case,  a non-dilute picture is much harder to handle and introduces considerable
complications into the semiclassical roadmap. The difficulty is not
necessarily an impossibility. For example,  the space of
self-dual configurations contains structures with all sort of
overlapping and non-diluteness, but still can be parametrised in terms
of a few parameters known as ADHM data~\cite{Atiyah:1978ri}. The dilute instanton pattern 
is included as a corner in this much richer configuration space. Other
self-dual configurations look very different. In particular, some can be
described by an ensemble of  lumps  of fractional topological charge 1/N. 
These other configurations not only lead to a non-vanishing
topological susceptibility, but also to a non-zero value of the string
tension. These and other attractive properties motivated  
a proposal for the structure of the QCD vacuum by our group
based on fractional topological charge structures~\cite{tonyfrac}. 
The proposal is not devoid of problems since  a realistic description
of the vacuum should allow the simultaneous presence of topological objects of
both signs, which are not classical solutions. 

Lattice Gauge theory provides a first-principles framework to study
non-perturbative aspects of the theory. It has been very successful 
in supporting QCD as the basic theory underlying strong interactions. 
Furthermore, it has enabled precision tests of the predictions of the
Standard Model, a necessary ingredient for detecting  evidence for any 
extension~\cite{latticeconf}. However, there are still many aspects 
which remain to be understood, some of which play a major role for 
candidate technicolor theories, for example. A priori, the lattice 
approach can also help in  solving the questions referring to the validity
of the semiclassical approach and the elucidation of whatever classical 
structures can account for the properties of the vacuum.
Unfortunately, the general statements made at the beginning of this
section imply that  Monte Carlo generated lattice configurations contain a
high level of  stochastic short-wavelength noise. This noise completely 
masks the long-wavelength topological structure information.
This justifies the necessity for  finding a procedure that would enable 
one to reduce this noise and  surface the underlying long-wavelength 
structures present in them.

A {\em filtering method} is a procedure  which takes as input a gauge field
configuration with stochastic noise and produces a new one which
preserves its long-range and topological structure and
substantially reduces the short-range stochastic noise. 
Many are the methods that have been explored to achieve this purpose.
They range from those which iteratively smooth the gauge field, like
smearing~\cite{Albanese:1987ds}-\cite{Morningstar:2003gk}, 
cooling~\cite{Teper:1985rb}-\cite{GarciaPerez:1998ru}, or the  
Wilson flow~\cite{Luscher:2010iy}, to those that attempt at 
reproducing the topological density through a truncation of the 
spectral density of operators like the Dirac~\cite{Hands:1990wc}-\cite{Moran:2010rn}, 
or the Laplacian operators~\cite{Bruckmann:2005hy}.
Ideally, any such method should produce a clean picture of the underlying 
long-range structures without introducing 
biases and distortions of the final data. However, all of the methods mentioned above
have been criticized in this respect since the amount of filtering they introduce depends
on ad-hoc parameters such as the number of cooling or smearing steps, 
or the cut-off at which the filtering operators are truncated.
Still, they have been used quite successfully in analyzing global as well as local
properties of the QCD topological structures. Recent reviews of the results obtained
employing these methods as well as comparison of the performance of different filtering 
procedures can be found i.e. in  Refs.~\cite{Solbrig:2007nr}-\cite{Bruckmann:2009vb}.
The aim of the present work is to explore and develop a particular implementation  
of a filtering method, introduced in Ref.~\cite{Gonzalez-Arroyo2006}, which does not suffer from
the ambiguities mentioned above. The basic point which motivated
the proposal is the idea that an optimal filtering method should leave smooth 
classical configurations invariant. Note that this is a non-trivial requirement,  
not satisfied for instance by any finite truncation of the spectral density 
of the Dirac operator in the fundamental representation. 
The key point exploited in Ref.~\cite{Gonzalez-Arroyo2006}
is that Supersymmetry equates the density of certain zero-modes of the 
Dirac operator in the adjoint representation with the gauge field action 
density~\cite{Novikov:1985ic}. However, in the
presence of noise both observables behave in a very different fashion. 
On general grounds it is to be expected that the density of the supersymmetric
zero-modes tends to suppress the effect of high frequency noise as compared to the 
action density itself, thus providing a filtered image of the latter. 
In Ref.~\cite{Gonzalez-Arroyo2006} the idea was presented 
and a series of tests were carried on to show that the method works satisfactorily.

In this paper we take this idea and develop one of its most elegant
and well defined implementations. This can be translated to the lattice by
making use of the overlap Dirac operator~\cite{Neuberger:1997fp},
which shares many nice features with its continuum counterpart. Preliminary results 
have been presented  in~\cite{latproc}-\cite{alfonsothesis}.  The structure of
the paper is as follows. In the following section we explain the
definition of the filtering procedure in its continuum version
and discuss why it is supposed to produce a filtered image of the gauge 
action density even in the presence of noise. Section~\ref{s.num}
provides the details of the lattice implementation of the procedure 
in terms  of the overlap Dirac operator.
In section~\ref{s.smooth}  we carry on a systematic study 
of how the filtering affects smooth configurations, both 
solutions of the classical equations of motion and others that are not, such as 
instanton - anti-instanton pairs. A basic
goal is to quantify the distortions and biases that the method could
introduce.  The final test, presented in section~\ref{s.heat},
corresponds to the analysis of noisy configurations that
have been obtained by applying several heat-bath Monte Carlo sweeps to 
a smooth instanton configuration. We will show that the method spectacularly 
succeeds in filtering away the ultraviolet noise. 

Having tested the method, we embark in section~\ref{s.dynamics}
in a possible application which is novel and might have important implications in hot 
problems such as the observed slow thermalization of Monte Carlo 
updates at high values of the coupling~\cite{critslow}-\cite{critslow3}. 
Starting with a classical configuration we apply a series of heat-bath steps at several values
of the coupling. This introduces stochastic noise which, even for very
small number of heat-bath steps, masks the classical configuration from which we
started. The updating procedure also induces  a motion along the space 
of classical configurations, changing its  moduli parameters. In
particular, for instantons, which have been our main focus, the
position and size of the instanton performs a stochastic motion. Our 
method can be used to study this motion and test if it follows the 
theoretical expectations. Results are quite impressive and might allow 
the numerical determination of the relative quantum weights of other
families of classical configurations. 

The results and conclusions of the paper are summarised in the last section.

\section{Construction of the filtering method}
\label{s.main}
As explained in the introduction, the main idea underlying the method is the 
relation between gauge field structure and zero-mode density implied 
by Supersymmetry. More precisely, we make use of the fact that given 
a solution of the classical equations of motion, the Dirac operator in the
adjoint representation has 
a zero-mode, called {\em the supersymmetric zero-mode}, whose density 
coincides with the action density~\cite{Novikov:1985ic}. 
As a matter of fact, decomposing 
the mode into its two chiral components, the corresponding densities 
follow the shape of the self-dual and anti-self-dual parts of the
action density. The general details are given in Ref.~\cite{Gonzalez-Arroyo2006} 
and will be explained below. For the sake of clarity  we will start by describing 
the method for the particular implementation which we are using in this
paper. 

Given a gauge field configuration $A_\mu(x)$, we can construct the  
corresponding Weyl operators $\bar D$ and $D$ for the adjoint 
representation. The Weyl $D$  operator is given by $D=D_\mu \sigma_\mu$, with
$\sigma_\mu=(\II, -\imath \vec{\tau})$ and    $\vec{\tau}$ the Pauli
matrices. The other operator $\bar D=D_\mu \bsig_\mu$, is minus the
adjoint of $D$. The Weyl matrices $\sigma_\mu$ and 
$\bsig_\mu$, satisfy the relations 
\be
\sig_\mu \bsig_\nu  = \eta^{\mu \nu}_\alpha \bsig_\alpha\quad ; \quad \quad
\bsig_\mu \sig_\nu  = \bar{\eta}^{\mu \nu}_\alpha \sig_\alpha \quad ,
\label{eq.etas}
\ee
where $\eta^{\mu \nu}_\alpha$ and $\bar{\eta}^{\mu \nu}_\alpha$ are
the `t Hooft symbols (summation over repeated indices implied).  

Using the Weyl operators we can construct the following hermitian 
positive  operator:
\be
-D \bar D=  -D_\mu D_\nu \sigma_\mu \bar \sigma_\nu= -\eta^{\mu
\nu}_\alpha \bar \sigma_\alpha D_\mu D_\nu \quad . 
\ee
The spectrum of this operator is doubly degenerate. This follows from 
the reality of the covariant derivative in the adjoint representation. 
Given an eigenvector $\psi$, one can construct an orthogonal  one of the 
same eigenvalue  as $\sigma_2 \psi^*$. The two bi-spinors can be
combined into a single quaternionic eigenvector $\Psi= \Psi_\mu \sigma_\mu \equiv (
\psi, \sigma_2 \psi^*)$.  
We are identifying the space of quaternions with a real subspace of the space of
$2\times 2$ complex matrices. The Weyl matrices $\sigma_\mu$
are a basis of the quaternionic space. Using this basis the operator  
can be seen as a  $4\times 4$ matrix whose components can be obtained by 
making use of Eq.~(\ref{eq.etas}). 
Having presented the basic terminology, we proceed with the explanation of the method.

The basic object, in terms of which the 
filtering procedure is formulated, is $\Op \equiv -PD\bar D P$, where  $P$ denotes 
the projector onto the 3 dimensional subspace generated by the Pauli matrices. 
In components it is given by:
\be
O^{(+)}_{j k}=  -D_\mu D_\nu \eta^{\mu \nu}_\alpha \bar \eta^{\alpha
k}_j = -\delta_{ j k} D_\mu^2 - \epsilon_{ i j k} \eta^{\mu
\nu}_i D_\mu D_\nu \quad .
\ee
The right hand side can be processed using the
expression for the commutator of covariant derivatives:
\be
O^{(+)}_{j k}= -\delta_{ j k} D_\mu^2 +\frac{i}{2}  \epsilon_{ i j k} \eta^{\mu
\nu}_i F^{\mu \nu} = -\delta_{ j k} D_\mu^2 + i  \epsilon_{ i
j k} (E_i+B_i) \quad .
\label{eq:oplus}
\ee
This restriction preserves the real, hermitian,  positive  semi-definite 
character of the $-D \bar D$ operator. 

The filtered image of the self-dual part of the action density is defined as
the density of the eigenmode $\Psi^{(+)}_j(x)$ of lowest eigenvalue of the 
operator $O^{(+)}_{j k}$:
\be
\sum_{k=1}^3 O^{(+)}_{j k} \Psi^{(+)}_k(x) =\lambda_0^+
\Psi^{(+)}_j(x) \quad .
\ee
Indeed the components in colour space $\Psi^{a (+)}_j(x)$ are real
numbers, and the density is given by 
\be
\sum_{j=1}^3 \sum_{a=1}^{N^2-1} (\Psi^{a (+)}_j(x))^2 \quad .
\ee

The whole construction can be repeated for the other chirality. The
main difference is that the $\eta$ and $\bar{\eta}$ symbols are exchanged.
Hence the final formula becomes 
\be
O^{(-)}_{j k}= -\delta_{ j k} D_\mu^2 -\frac{i}{2}  \epsilon_{ i j
k} \bar \eta^{\mu
\nu}_i F^{\mu \nu} = -\delta_{ j k} D_\mu^2 - i  \epsilon_{ i
j k} (E_i-B_i) \quad .
\label{eq:ominus}
\ee
The corresponding mode $\Psi^{(-)}_j(x)$ is the filtered image of
the anti-self-dual part of the gauge field. 
For the ease of notation we will be referring to the lowest eigenvectors 
of the $\Opm$ operators as the adjoint filtering modes or AFM modes.

Having presented the method, we will explain the reasons why it is 
supposed to produce a filtered image of the gauge field configuration. 
To start with, we will focus on smooth configurations which are
solutions of the classical equations of motion. In this case the 
filtered image should produce an undistorted image of the
configuration itself. This is precisely the role played by
Supersymmetry and the supersymmetric zero-mode. Given a solution 
of the euclidean classical equations of motion, we can decompose 
its field strength into its selfdual and
anti-selfdual parts, as follows:
$$ F^{\mu \nu}= \eta^{\mu \nu}_i \chi_i + \bar \eta^{\mu \nu}_i \chi_i' \quad .$$
Combining the equations of motion with the Bianchi identities one sees 
that the  self-dual part satisfies the equation: 
\be
 \eta^{\mu \nu}_i D_\mu \chi_i=0 \quad ,
\ee
and a similar equation, with $\eta$ replaced by $\bar \eta$, for the anti-selfdual 
part. Introducing a quaternionic field $\Psi(x)=\sigma_i \chi_i(x)$, one can show 
(using the properties of the `t Hooft $\eta$ symbols) that the previous equation
can be written as 
\be 
\bar D \Psi(x)=0 \quad .
\ee 
This equation expresses that $\Psi(x)$ is a  zero-mode of the
Dirac equation in the adjoint representation. This is the
supersymmetric zero-mode, whose density is exactly proportional to
that of the self-dual part of the field. Obviously, $\Psi(x)$ is also
a zero-mode of the positive hermitian $-D \bar D$ operator. 
Noticing that $\Psi(x)$ is a linear combination of the Pauli matrices 
alone, one concludes that $\chi_i =(E_i+B_i)/2$  is also a zero-mode of the
operator $O^{(+)}$. Now, one understands better the role played by 
the projection onto the three-dimensional subspace leading to the
construction of $O^{(+)}$. Indeed, the Weyl operator $\bar D$ might have many
zero-modes. This is necessarily the case for classical solutions
having positive definite topological charge (number of adjoint
zero-modes $\ge 2NQ$). However, the supersymmetric zero-mode is indeed 
unique in that its quaternionic field is traceless, at all points of
space and for all colour components. In other words it is a purely
imaginary quaternionic field. This property, called {\em reality
condition} in Ref.~\cite{Gonzalez-Arroyo2006}, lies at the heart of our method. 
We lack a
general proof of uniqueness, but in the particular case of 
self-dual gauge fields, the ADHM construction allows us to give formal
expressions for all the zero-modes. In quaternionic notation they 
can be written as $\chi_\alpha \sigma_\alpha$. The reality
condition amounts to demanding $\chi_0(x)=0$. This, except for some
limiting cases, indeed singles out the supersymmetric zero-mode. 
Our approach in this paper has been to directly impose the reality condition
by looking at the projected operators $\Opm$. Alternatively, one could start by
exploring the low-lying spectra of the $-D\bar D$ operator, imposing a posteriori the
reality condition to select the supersymmetric zero-mode. This is the practical 
approach implemented in Ref.~\cite{Gonzalez-Arroyo2006}. Although
the results obtained were quite satisfactory, it requires
dealing with the multiplicity of the $-D\bar D$ zero-modes 
and turns out to be less efficient and numerically more costly than the present
proposal.

What do we know about the excited states of the spectrum of the 
$O^{(+)}$ operator. In general, for a finite volume, we expect a gap to develop in the
spectrum. The gap goes to zero for large volumes, but the
corresponding limiting states become non-normalizable. The operator has 
a single point discrete spectrum and a gapless continuum spectrum.
Furthermore, the eigenmodes would look rather structureless 
and easily distinguishable from the supersymmetric zero-mode. We will see
how these expectations are realized in our numerical results.

Certainly a possible analytical testing ground is that of a single SU(2)
instanton.  Here the only scale is clearly $\rho$, so that we might use it as a
unit of lengths and energies.  The operator $\Op$ can be rewritten using the 
formula of the covariant Laplacian for an instanton in SU(2) \cite{'tHooft:1976fv} 
\be
O^{(+)}=  -\mathbf{I} \left( \left(\frac{\partial }{\partial
r}\right)^2 +\frac{3}{r} \frac{\partial}{\partial r} -\frac{4}{r^2}
\vec{L_1}^2-\frac{8}{r^2+1} \vec{T}\cdot \vec{L}_1 - \frac{4
r^2}{(r^2+1)^2} \vec{T}^2\right)    +\frac{8}{(r^2+1)^2}  \vec{T}\cdot
\vec{T'} \quad .
\ee
In the previous formula:
$$ 
L_1^a = -\frac{i}{2} \eta_a^{\mu \nu} x_\mu \frac{\partial} {\partial x_\nu}\quad ,
$$ 
and $ \vec L_1^2= -(x_\mu \partial_\nu - x_\nu \partial_\mu)^2/8$, 
where summation over repeated indices is understood.
In addition $\vec{T}$ and $\vec{T'}$ are the generators of SU(2) in the
adjoint representation acting respectively on the colour or on the spinorial 
indices. 

If we restrict to purely radial eigenvectors and split upon
eigenstates of the direct product representation $\vec{T}+
\vec{T'}$,
with isospins $t=0,1,2$, we get:
\be
O^{(+)}=  -  \left(\frac{\partial }{\partial
r}\right)^2 -\frac{3}{r} \frac{\partial}{\partial r}  + \frac{8
}{(r^2+1)}     + \frac{8}{(r^2+1)^2}  \frac{t(t+1)-6}{2}  \quad .
\ee
We might write the rotationally invariant eigenvectors  as:
\be
\Psi= \frac{G(r^2)}{(r^2+1)^2}\quad .
\ee
The eigenvalue equation then reduces to
\be
- x G''(x) +2 \frac{x-1}{x+1} G'(x) + \frac{t(t+1)}{(x+1)^2}G
=\frac{\lambda}{4} G(x) \quad .
\ee
Obviously for $t=0$,  $G(x)=1$ is a solution of vanishing eigenvalue.
This is the expected supersymmetric zero-mode. 
It is easy to see that there are in addition 5 non-normalizable zero-modes 
for $t=2$ and $G(x)= (r^2+1)^3$.

Up to now, we have just analysed the behaviour of the filtering
procedure for solutions of the classical equations of motion, for which 
the filtering mode is just the supersymmetric zero-mode. For other 
smooth configurations the lowest eigenvalue will not be zero. Indeed,
if we take the self-dual part of the gauge field
$\chi_i(x)=(E_i(x)+B_i(x))/2$ and apply the operator to it, we get 
\be
O^{(+)}_{j k} \chi_k(x)= -\frac{1}{2} \eta^{\rho \nu}_j D_\rho j^\nu \quad ,
\ee
which only vanishes if the current, $j^\nu =  D_\mu F^{\mu \nu}$, vanishes. 
From here we get an upper bound on the minimal eigenvalue of $O^{(+)}$:
\be
\lambda_0^+ \le  \frac{\int dx \mathrm{Tr}(j^\nu(x)
j^\nu(x))}{4 \int dx \mathrm{Tr}(\chi_k(x) \chi_k(x))} \quad ,
\label{eq.lmin}
\ee
If the smooth configuration is made of local structures which are 
approximate solutions of the classical equations of motion, the current 
would be very small in those regions and the filtered mode
$\Psi^{(\pm)}_k(x)$ will be close to 
the self-dual action density. The limit to this situation would be the
region in which level crossing appears. We will explore this situation in our
numerical analysis, to be presented later.

In order to understand the behaviour of the $O^{(\pm)}$ operators for
non-classical configurations, one could study small perturbations about the
classical solutions: $A_\mu+\epsilon \delta A_\mu$. To derive the
appropriate formula it is better to regard $O^{(+)}$ as the projected
version of $-D \bar D$. The eigenvalue equation, to leading order in
$\epsilon$, takes the form 
\be
-P D\delta \bar D \Psi - P D\bar D \delta \Psi = \delta \lambda \Psi \quad .
\ee
Projecting onto $\Psi$ one realizes that $\delta \lambda$ vanishes
(hence it is order $\epsilon^2$). The equation for the eigenvector is
\be
P D\bar D \delta \Psi = - P D\delta \bar D \Psi  \quad .
\ee
The right-hand side can be evaluated to 
\be 
i \sigma_\nu \bar \sigma_\rho \sigma_j [D_\nu \delta A_\rho, \Psi_j]
+ 2 i \sigma_j [\delta A_\rho , D_\rho \Psi_j ] \quad . 
\ee
Now, without loss of generality we might impose the background gauge condition 
on $\delta A_\rho$. With this we can rewrite the right-hand side as 
\be
i \bar \sigma_k \sigma_j [ \delta E_k +\delta B_k, \Psi_j]+
2 i \sigma_j [\delta A_\rho , D_\rho \Psi_j]\quad .
\ee

After several operations one arrives at the equation 
\be
 -D_\mu^2 \delta \Psi_j +i \epsilon_{i j k} [E_i+B_i, \delta \Psi_k]=
-i \epsilon_{i j k} [\delta E_i+\delta  B_i, \Psi_k] - 2 i [ \delta
A_\mu, D_\mu \Psi_j] \quad ,
\label{perturb}
\ee
where $\Psi_j=(E_j+B_j)/2$. 

To see the filtering properties of the method, we might consider a deformation $\delta A_\mu $
involving  high frequencies. For that purpose we make $\delta A_\mu $
proportional  to $e^{i px}$ for high values of $p$.  The corresponding 
field strengths $\delta E_i$, $\delta  B_i$  will then be proportional to
$pe^{i px}$. Thus, in the corresponding action density the noise terms 
are amplified with a factor $p^2$. Let us now look at the expected
behaviour of the filtered mode $\delta \Psi_k$.  The right-hand side
of Eq.~(\ref{perturb})  will also contain terms proportional to $p e^{ipx}$. 
To solve the equation we should decompose this term into eigenstates of 
the $O^{(+)}$ operator. The decomposition  should be dominated by states of
high eigenvalue, which are essentially given by plane waves $e^{i
qx}$. Since, the field-strength of the zero-mode is smooth, the
typical wavelengths involved are $q\approx p$. The same reasoning
tells us that the corresponding eigenvalue of the $O^{(+)}$ operator
is $q^2$. Applying, these considerations to Eq.~(\ref{perturb}) we
conclude that $\delta \Psi$ goes like $p/p^2=1/p$. The new filtered
mode  density  has a $1/p$ suppression of the added noise. 
This is precisely the desired effect, which makes our 
method a filtering procedure for the gauge field configurations.

\subsection{Lattice implementation}
\label{s.num}

There are two important ingredients involved in the construction of
the $O^{(\pm)}$ operators: chirality and the reality of the covariant
derivative. In implementing the method on the lattice it is convenient 
to preserve these properties. This can be achieved by using the
overlap operator \cite{Neuberger:1997fp} as lattice
Dirac operator. This will be explained below. Some results obtained with 
this procedure were presented in Ref.~\cite{Perez:2010ja}.

The lattice version of the $O^{(\pm)}$ operator is given by 
\be
\label{latticeops}
O_L^{(\pm)} = P P_\pm  (\gamma_5 D^A_{ov})^2 P_\pm P \quad ,
\ee
where $P_{\pm} = \frac{1}{2}(1\pm\gamma_5)$ and $P$ is the projection
onto the space of purely imaginary quaternions as before. The 
overlap Dirac operator is given by the formula:
\begin{equation}
 D^A_{ov} = \frac{M}{a} \left(\mathbf{I} + \gamma_5 \boldsymbol{\epsilon}\right) \quad ,
\end{equation}
where $\boldsymbol{\epsilon}$ is a hermitian, unitary matrix given by:
\be
\boldsymbol{\epsilon}= \frac{H}{\sqrt{H^2}} \quad ,
\label{eq.epsmatrix}
\ee
and $H = \gamma_5(-M + a D_{WD}^A)$. 
The matrix $D_{WD}^A$ is the standard massless Wilson-Dirac 
operator in the adjoint representation given by:
\be
a D_{WD}^A(n,m)= -\frac{1}{2} \sum_{\mu} \left( (\mathbf{I}-\gamma_\mu) U_{\mu}^A(n)
\delta_{m\, n+\hat{\mu}} + (\mathbf{I}+\gamma_\mu) U^{A\, \dagger}_{\mu}(m)
\delta_{m\, n-\hat{\mu}} -2 \mathbf{I} \delta_{n\, m}\right) \quad , 
\ee
where $n$ and $m$ label points in the 4-dimensional lattice of spacing
$a$, and $\hat{\mu}$ is the unit vector in the $\mu$ direction. 
The gauge field links in the adjoint representation can be obtained in
terms of links in the fundamental as follows:
\begin{equation}
 \left(U_{\mu}^A\right)_{ab} = 2 \mathrm{Tr}\left(U^\dagger_\mu T_a U_{\mu} T_b \right) \quad ,
\end{equation}
where $T_a$ are the generators of SU(N) in the fundamental
representation with the standard normalization $\mathrm{Tr}(T_a
T_b)=\frac{1}{2}\delta_{a b}$.  Notice that the links in the adjoint
representation are  real orthogonal matrices. 

The overlap Dirac operator is known to satisfy $\gamma_5$-hermiticity
and the Ginsparg-Wilson relation 
\begin{equation}
 \left\{\gamma_5,D^A_{ov}\right\} = \frac{a}{M} D^A_{ov}\gamma_5
 D^A_{ov} \quad .
 \end{equation}
 These two conditions imply that $(\gamma_5 D^A_{ov})^2$ commutes with
 $\gamma_5$. The positive and negative chirality components are the
 lattice counterpart of $-D\bar{D}$ and $-\bar{D}D$. 
 
Finally, we should verify that the projection onto the purely
imaginary quaternions operates in the same way as in the continuum.  
The reality of the gauge field links in the adjoint representation
implies the double-degeneracy of the eigenvalues of 
$H$. This follows from the equation 
\be
\label{chargeconj}
H^*= \widetilde{C}^{-1}H \widetilde{C} \quad ,
\ee
where $\widetilde{C}$ is the charge conjugation matrix satisfying
$\gamma_\mu^*=\widetilde{C}^{-1}\gamma_\mu \widetilde{C}$. Eq.~(\ref{chargeconj})
 also holds for the overlap Dirac operator and for $(\gamma_5 D_{ov})^2$. In the
Weyl representation of the Dirac matrices
\begin{equation}
 \gamma_\mu = \left(\begin{array}{cc} 0 & \sigma_\mu \\
 \bar{\sigma}_\mu & 0\end{array}\right) \quad , 
 \end{equation}
the matrix $\widetilde{C}$ can be chosen as  $\gamma_5\gamma_0\gamma_2$.
Thus, following the same procedure as in the continuum, the chiral 
components  of  $(\gamma_5 D_{ov})^2$ can be seen as a matrix acting
on quaternions. Restriction to the subspace of purely imaginary
quaternions leads to the  $O_L^{(\pm)}$ matrices defined in
Eq.~(\ref{latticeops}).

In our lattice implementation of the overlap operator, we have used the rational expansion 
for the $\boldsymbol\epsilon$ matrix proposed in
\cite{Neuberger:1998my,Edwards:1998yw}:
\begin{equation}
 \boldsymbol{\epsilon} = \lim_{n \rightarrow \infty} \frac{H}{n}\sum_{s=1}^{n}\frac{\omega_s}{H^2 + \sigma_s} \quad ,
\end{equation}
where $\omega_s =\cos^{-2}(\frac{\pi}{2n}(s-\frac{1}{2}))$ 
and $\sigma_s = \tan^2(\frac{\pi}{2n}(s-\frac{1}{2}))$. 
The action of the $(H^2 + \sigma_s)^{-1}$ matrices over a vector $x$ is computed by using the
method of multiple shifts described in \cite{Jegerlehner:1996pm}. 
As a compromise between performance and stability we have taken $n = 60$ and the 
$\textit{tolerance}$ parameter that controls the conjugate gradient accuracy is taken to $10^{-9}$. 
In order to minimize the storage needs, we use the \textit{two-pass} algorithm proposed by Neuberger \cite{Neuberger:1998jk}.
The pseudo-code necessary for the implementation is well described in Ref.~\cite{Neuberger:1998jk}.

In order to compute the low-lying spectrum of the $\Opm_L$ operators we have used the 
conjugate gradient algorithm~\cite{Kalkreuter:1995mm}. 
The eigenvectors are computed with an accuracy such that:
\be
\frac{\langle \Psi, (\Opm_L-\lambda) \Psi \rangle}{\langle\Psi,  \Psi \rangle}  < \omega \quad ,
\ee
with $\omega= 10^{-8}M^2/a^2$. This guarantees no
appreciable errors in the eigenvector densities. The speed at which 
convergence is achieved depends very much on the gauge field configuration and on the
initial vector for the conjugate gradient algorithm. Typically, for the
$Q=1$ cases analysed in the paper, the computer time needed
to determine the lowest eigenvector grows as $N_s^4$, with  a proportionality 
constant that increases as $\beta$ decreases.
In some cases we have seen that a suitable choice of
the initial vector for the conjugate gradient can reduce this time by a factor
of 5.

To check that our results are not biased by the choice of
the $\boldsymbol{\epsilon}$ matrix expansion,
we have repeated the computation of the low-lying spectrum on some of the configurations with
the Lanczos algorithm proposed
in \cite{Borici:1998mr}. The results obtained by both methods are compatible within errors.

Finally, we comment on the choice of the Wilson-Dirac mass term $M$.
Most of our results were obtained for $M=1.4$. This value is a
compromise which allows the filtering method to  detect 
fairly small instantons 
without sacrificing performance~\cite{latproc,alfonsothesis}.
Nevertheless, we have done several tests to check the dependence
of our results upon the value of $M$. In all cases, including noisy
and smooth gauge field configurations, no significant changes are
observed on the density of the AFM. This contrasts with
the results in Ref.~\cite{Moran:2010rn} obtained using the modes of the Dirac
operator in the fundamental representation as filtering modes. In that case, the claim is
that the value of $M$ monitors the amount of filtering obtained for the configuration.

\section{Testing the method with smooth configurations}
\label{s.smooth}

In this section we will test the numerical implementation of the filtering method on a set of smooth
lattice configurations. We will focus on two generic situations. First, we will look at lattice 
configurations that correspond, in the continuum limit, to solutions of the 
classical equations of motion. 
This includes discretized $Q=1$ instantons and calorons \cite{kvanbaal1,lee1} as well as higher topological charge 
objects. The filtering method should work well in this case, but the analysis will allow to quantify 
discretization and finite volume effects. Second, we will analyze smooth configurations which 
are not solutions of the continuum equations of motion. We will focus in particular on the case
of instanton - anti-instanton pairs, testing the ability of the filtering method 
to detect structures in both chiral sectors. 
In all cases we will analyze the low lying spectrum of the $\Opm_L$ operators, 
paying particular 
attention to the occurrence of level-crossing that could disturb the identification 
of the adjoint filtering mode.

\subsection{Classical configurations}

The filtering operator is designed to select the supersymmetric zero-mode
if applied to classical backgrounds. This is guaranteed in the continuum
but, for the method to work, it is essential that this property is 
preserved when both the $O^{(\pm)}$ operators and 
the classical configurations are discretized. 
Lattice artefacts can be considered as a particular case of the deformations discussed
in section~\ref{s.main}. We expect them to induce an ${\cal O}(a^2)$ lifting
of the AFM eigenvalue and a small modification of its
density through mixing with the excited states. Although this contamination goes away 
in the continuum limit, it is important to quantify 
the distortions it may introduce in the procedure. 

In what follows we will analyze several $Q=1$ and $Q=2$, SU(2) configurations
discretized on the lattice. We will show that the method spectacularly succeeds 
in identifying classical lumps down to sizes of the order of two lattice spacings. 
Results will be presented for SU(2) instantons
but we have also checked that the filtering procedure works well for other classical 
solutions as finite temperature calorons~\cite{kvanbaal1,lee1}, or fractional 
charge instantons~\cite{'tHooft:1981sz,GarciaPerez:1989gt}. 

\subsubsection{Single instanton configurations}
\label{s.q1}

For the purpose of the test we have generated a set of $Q = 1$, SU(2) lattice instanton 
configurations for several lattice volumes ($N_s^4$) and two different sets of boundary 
conditions. In addition to a periodic torus (p.b.c.) 
we have also considered a torus with twisted boundary conditions (t.b.c.)~\cite{'tHooft:1979uj}, 
corresponding to $\vec m= (0,0,0)$ and $\vec k =(1,1,1)$. 
The reasons to employ twisted boundary conditions are twofold. To start with, the gap in the
spectrum of the filtering operator is strongly dependent on the boundary conditions.
It is easy to understand why this is the case by analysing the spectrum
on a trivial background. For t.b.c. the set of classical vacua is discrete
and the spectrum has a gap of size $\pi^2/L^2$, for  $A_\mu=0$
and finite box size $L\equiv N_s a$. Periodic boxes, on the contrary, have a continuum  
of classical vacua~\cite{GonzalezArroyo:1981vw,Coste:1985mn} which support a 
continuum of constant density
adjoint zero-modes. On the background of instanton configurations,
these vacuum zero-modes become quasi-zero-modes for $\rho<< L$ and
can give rise to unwanted level-crossing in the spectrum. 
This situation is avoided in the twisted case, where the 
$\pi^2/L^2$ gap in the spectrum pushes any level-crossing to very large values of $L$.
There is a second reason to select twisted instead of periodic boundary conditions 
for the analysis of $Q=1$ instantons.
On a finite periodic box there is a restriction over the allowed topological charge 
of classical solutions, and $Q=1$ is not feasible~\cite{Braam1989}.
Although this restriction disappears in the thermodynamic limit, it gives rise to finite size effects
that can shift considerably away from zero the lowest eigenvalue of the 
filtering operator.
This limitation disappears, however, as soon as one imposes a non-zero twist.

\begin{figure}
\centerline{
\psfig{file=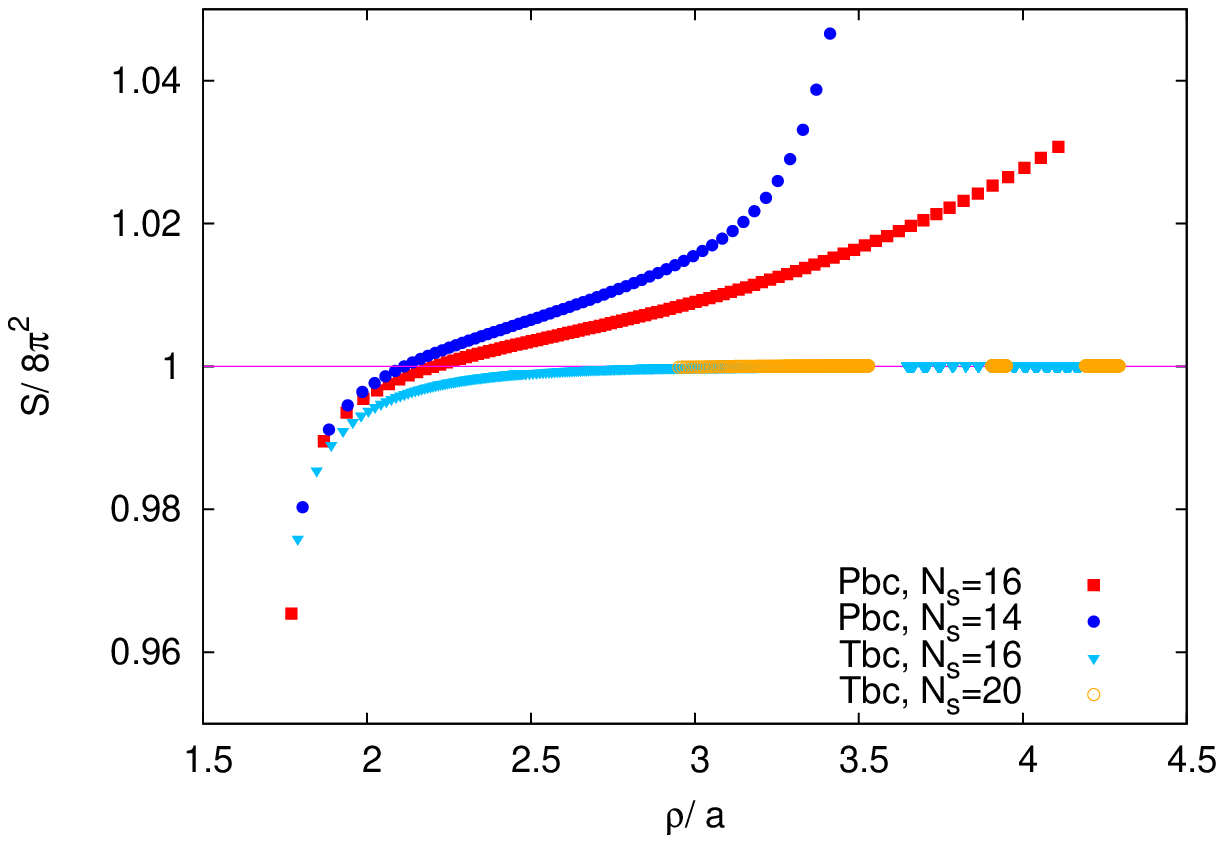,angle=0,width=0.48\textwidth}
\psfig{file=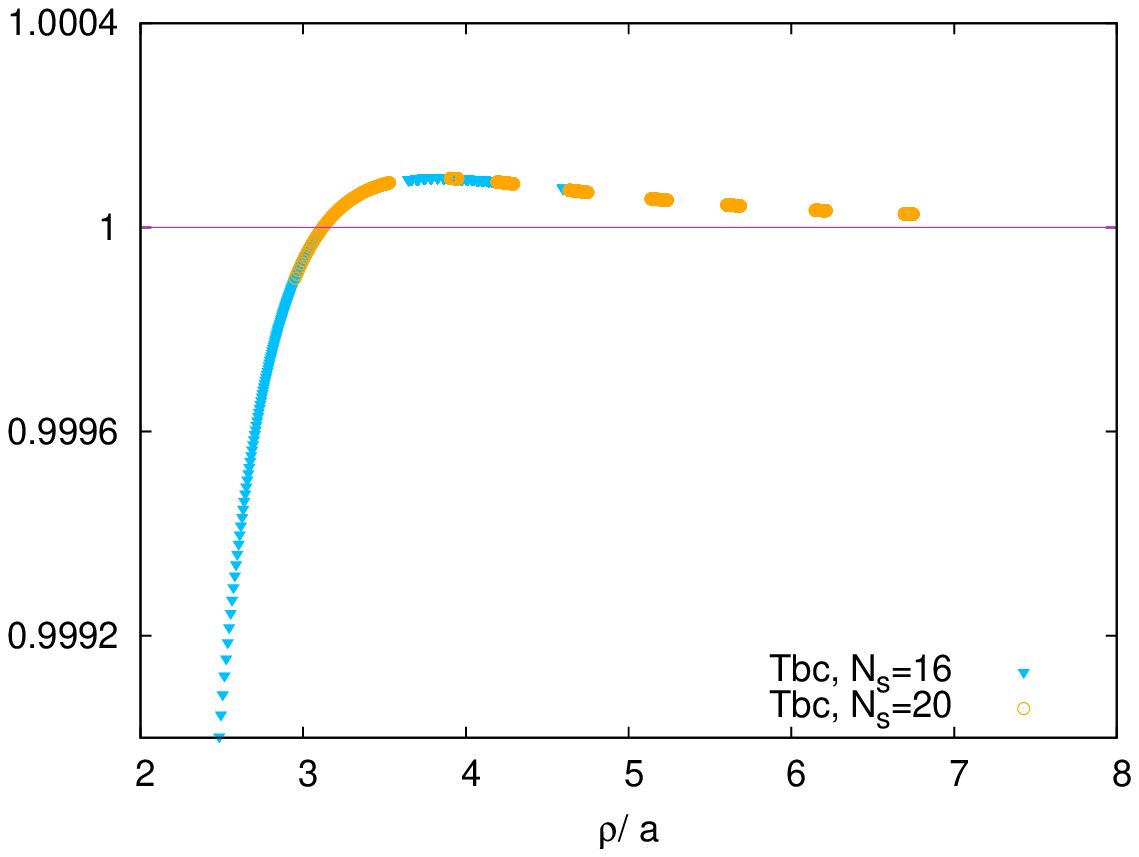,angle=0,width=0.48\textwidth}}
\caption{We display the dependence of the instanton action  (in units of $8\pi^2$)
versus the instanton size (in lattice units)
for several values of the lattice volume: $N_s=14$, 16, and 20.
(Left) Comparison of twisted and periodic boundary conditions; (Right) A blow up of the
plot for twisted boundary conditions (note the change of scale).
The action corresponds to the discretized $\epsilon=0$ lattice action.}
\label{fig:svrho}
\end{figure}

\begin{figure}
\centerline{
\psfig{file=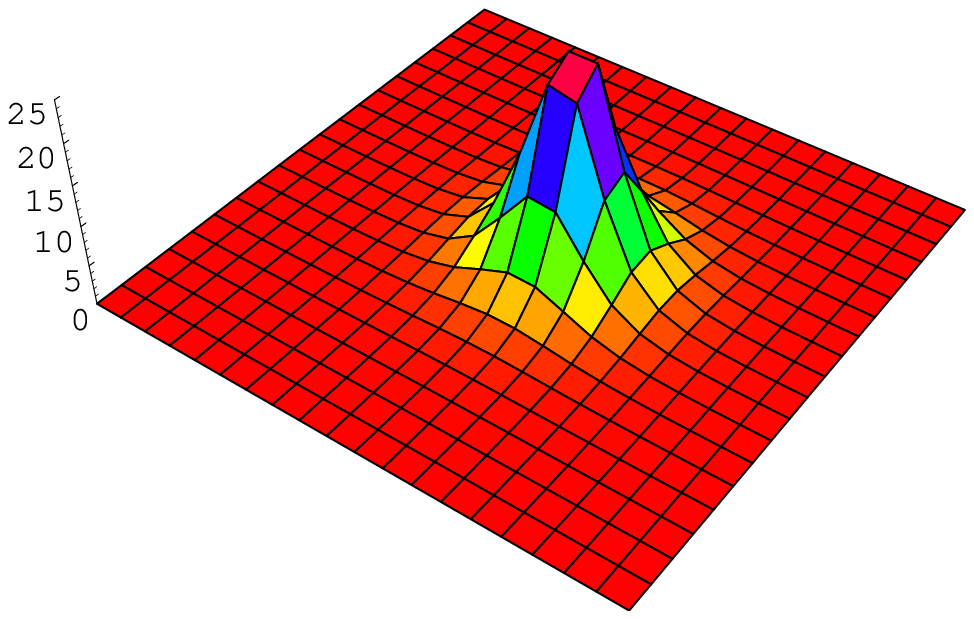,angle=0,width=0.5\textwidth}
\psfig{file=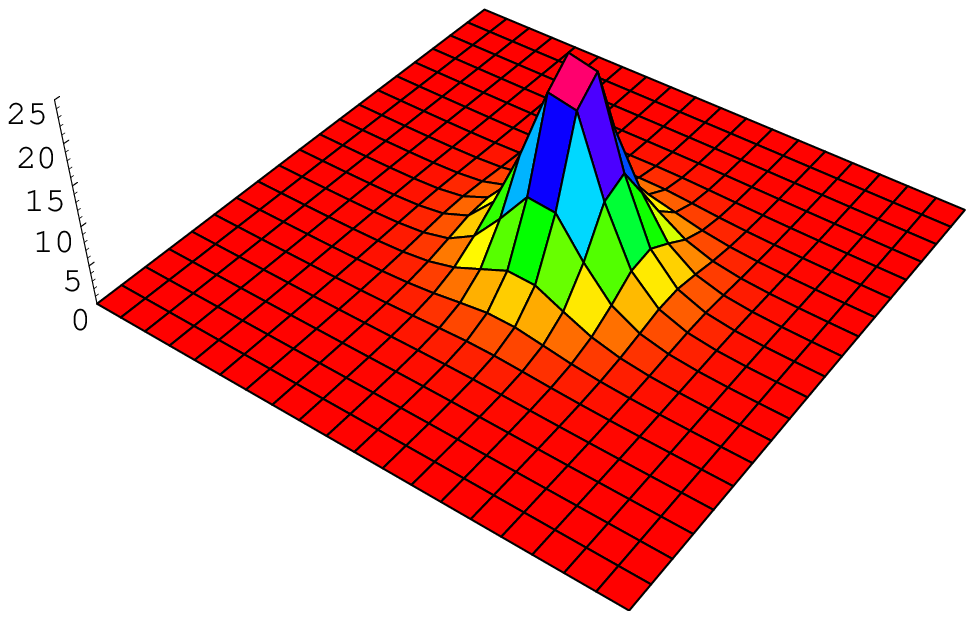,angle=0,width=0.5\textwidth}
}
\centerline{
\psfig{file=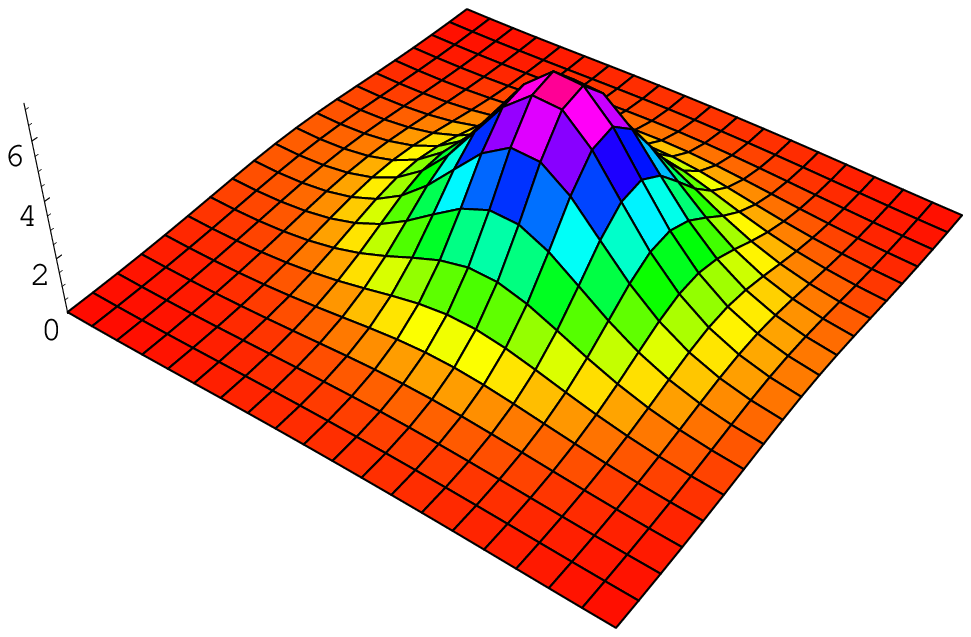,angle=0,width=0.5\textwidth}
\psfig{file=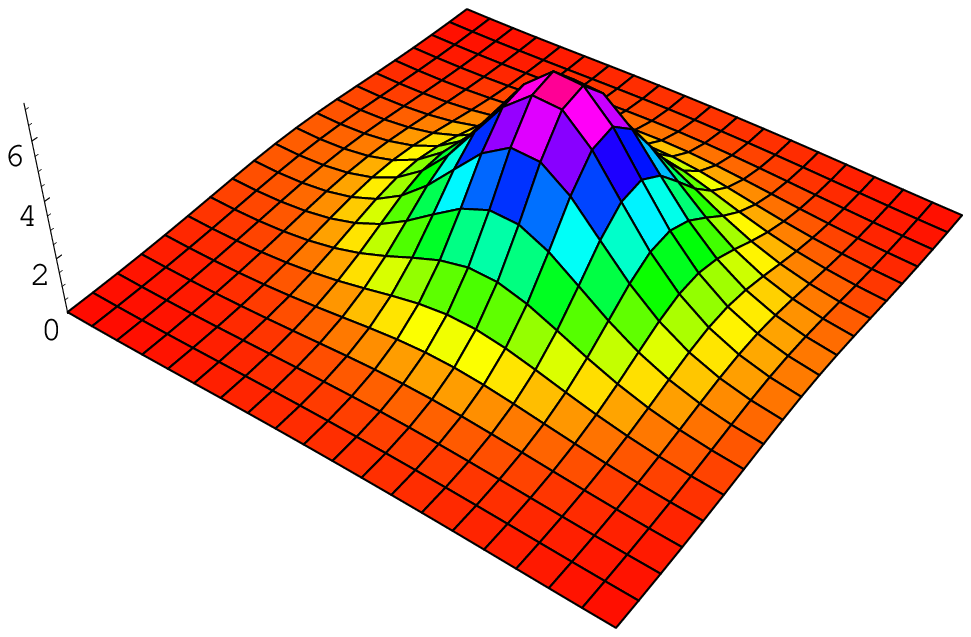,angle=0,width=0.5\textwidth}
}
\caption{Left: the self-dual part of the action density (in units of 
$8 \pi^2$), and  Right: The density of the lowest
eigenvector of the $\Op_L$ operator, both integrated over the two orthogonal directions.
The top and bottom plots correspond respectively to $Q=1$ instanton configurations of
sizes $\rho= 3 \  a$ and $\rho= 6.2 \ a$. All the data are for twisted boundary 
conditions and $N_s=20$.}
\label{fig:sdensq1}
\end{figure}

The results we will present correspond to instanton configurations of various 
sizes generated 
by cooling, an iterative update process that monotonically decreases the lattice 
action~\cite{Teper:1985rb}. Since the lattice breaks the scale
invariance, the fate of instantons 
under cooling depends crucially on the choice of lattice action employed in the cooling 
algorithm~\cite{Garcia Perez:1993ki}. 
One can take advantage of this and use cooling to modify at will the
instanton size parameter $\rho$. 
In this paper we have made use of the 
$\epsilon$-cooling method proposed in Ref.~\cite{Garcia Perez:1993ki}.  
It employs a variant of the Wilson action containing $1 \times 1$ and $2 \times 2$ 
plaquettes with 
relative coefficients tuned to control the ${\cal O}( a^2)$ scale violations of the lattice action.
In particular, $\epsilon=0$ is the choice that eliminates these corrections. 
Figure~\ref{fig:svrho} displays the dependence of the integrated $\epsilon=0$
instanton action versus the instanton size for several lattice volumes. For t.b.c. the 
deviation from the continuum instanton action amounts to less than one part per mille 
for instantons of size larger than $\rho = 2.5 \, a$. In that range we expect very 
small distortions in the spectrum of the $\Op_L$ operator. The situation is not 
as good for periodic boundary conditions since, as we have just discussed,
there are no exact classical solutions with $Q=1$ on a finite periodic box.
For this reason the deviation from the continuum action is generically much 
larger than with t.b.c.. 
The results that will be presented in this section correspond to several 
representative twisted and periodic configurations selected among the ones displayed in these plots.

For all the configurations analyzed we have computed the four lowest eigenvalues in each 
chiral sector. The sector of negative chirality is structureless.  
For positive chirality, as expected, one of the eigenvalues is distinctively smaller than 
the others. Its eigenvector provides the adjoint filtering mode (AFM) which should 
correspond to the supersymmetric zero-mode whose density reproduces the instanton 
action density. A comparison between the density of the AFM and the self-dual part 
of the action density is displayed in figure~\ref{fig:sdensq1} for two 
instantons of sizes $\rho= 3\, a$ and $\rho= 6.2\, a$, with $N_s=20$ and twisted 
boundary conditions. The qualitative agreement is impressive. 
For a more quantitative comparison we give in table~\ref{table:q1chi} 
the values of the instanton parameters obtained from the densities of the AFM 
($X_{\AFM}$, $\rho_{\AFM}$) and the self-dual part of the gauge field strength ($X$, $\rho$). 
They have been extracted by fitting the 1-dimensional instanton profiles to the data, 
following the procedure described in the appendix. 
Differences in the extracted instanton locations are negligible.
As for the sizes, the differences remain much smaller than those associated to 
different fitting procedures - see figure~\ref{fig:sizes} in the appendix.

\begin{table}
\centerline{
\begin{tabular}{|c|c|c|c||c|c|}\hline \hline
$ \Delta X_{0}$  &  $\Delta X_{1}$  &  $\Delta X_{2}$&$ \Delta X_{3}$ &
 $\rho_{\AFM}$ & $\rho$ \\ \hline \hline
-0.0007&0.0004&-0.0005 &0.0008 &2.992 &2.977 \\ \hline
-0.0001&0.0001&0. &0.0001 &6.175 &6.190 \\ \hline
\end{tabular}}
\caption{
For two different size $Q=1$ instantons, on a $20^4$ twisted lattice, 
we compare the instanton parameters determined from the 1-d profiles
of the AFM density and the self-dual part of the action density.
The corresponding size parameters are given by: $\rho_{\AFM}$ and $\rho$.
The difference in the position of the instanton center coming from both determinations
is given by: $\Delta X_\mu $.
}
\label{table:q1chi}
\end{table}

\begin{figure}
\centerline{
\psfig{file=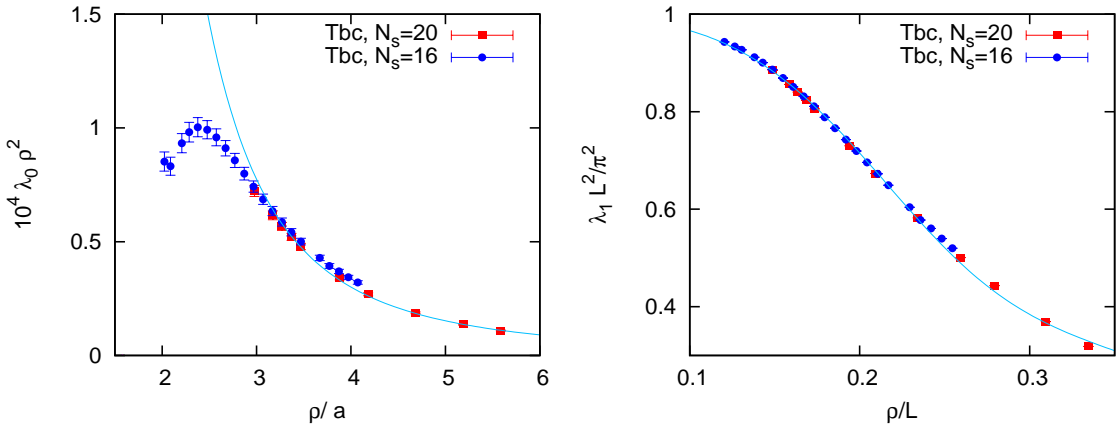,width=1.0\textwidth}
}
\caption{For a set of $Q=1$ twisted instanton configurations of size $\rho$,  
we display: Left: the  eigenvalue of the AFM,
as a function of $\rho/a$. The line corresponds to the fit in Eq.~(\ref{eq:l0_tw}).
Right: the first excited eigenvalue of the $\Opl$ operator,
as a function of $ \rho/L$. The line is the fit in Eq.~(\ref{eq:l1_tw}).}
\label{fig:itw_lowest}
\end{figure}

We proceed next to present a detailed analysis of the effect of the numerical implementation 
on the spectrum. It is important for the success of the method that no level-crossing takes 
place between the lowest eigenvalue and the excited states since this would induce a 
misidentification of the supersymmetric zero-mode.
We will discuss first the case of twisted boundary conditions which from this point of
view provides a much cleaner situation. 
As discussed before, the spectrum of the $\Opl$ operator is gapless in the infinite volume limit
with a continuum spectrum that corresponds to non-normalizable states. For finite volume however
there is a finite gap of order $\pi^2/L^2$ and no level-crossing is expected.
We will analyze in what follows how lattice artifacts modify this expectation,
and we will see that in the worst case situation level-crossing will only occur 
for very large lattice sizes $N_s > 500 $.
Figure~\ref{fig:itw_lowest} displays, for twisted boundary conditions and two different
lattice sizes, $N_s=16$ and 20, the lowest (left) and first
excited (right) eigenvalues of the $\Opl$ operator. Let us first focus on the analysis
of the lattice artifacts on the AFM eigenvalue ($\lambda_0$). In the plot it is displayed as a function of 
the instanton size in lattice units: $\rho/a$. The line drawn in the plot corresponds to a fit of the form:
\be 
\rho^2 \lambda_0 = \alpha \, \Big (\frac{a}{\rho}\Big )^2 + \beta \, \Big (\frac{a}{\rho}\Big )^4\quad ,
\label{eq:l0_tw}
\ee
with coefficients $\alpha = 2.0(2) 10^{-4}$, and $\beta= 44(2)  10^{-4}$. It describes well the data for 
$\rho > 3a$ and shows that the eigenvalue of the filtering mode approaches zero 
in the continuum limit, as expected. Results are given for instantons of size larger
than $\rho =2a$ to avoid problems in the convergence of the conjugate gradient algorithm.

\begin{figure}
\centerline{
\psfig{file=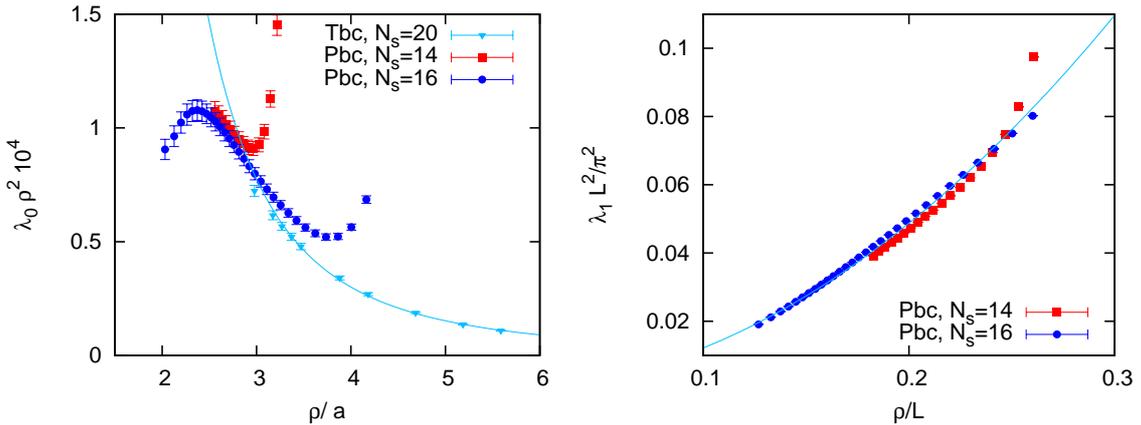,width=1.0\textwidth}
}
\caption{For a set of $Q=1$ periodic instanton configurations of size $\rho$, we display:
Left: the eigenvalue of the AFM,
as a function of $\rho/a$. The line corresponds to the fit in Eq.~(\ref{eq:l0_tw}).
For comparison we also include the data for twisted boundary conditions and $N_s=20$.
Right: the first excited eigenvalue of the $\Opl$ operator,
as a function of $ \rho/L$. The line is the fit in Eq.~(\ref{eq:l1_per}).}
\label{fig:iper_lowest}
\end{figure}

With respect to the first excited eigenvalue $(\lambda_1)$ a comparison between the $N_s=16$ and the 
$N_s=20$ results shows that it scales as $1/L^2$. The data can be fitted to 
\be
\lambda_1 = \frac{\pi^2}{L^2} \, \Big (\,  \half + \half e^{-b \, (\frac{\pi \rho}{ L})^4} 
- c \, \frac{\pi^2 \rho^2}{ L^2}\Big ) \quad ,
\label{eq:l1_tw}
\ee
where  $b=3.8(1)$, and $c= 0.16(1)$. From the fits to the two eigenvalues we can extract the ratio 
$\lambda_0 / \lambda_1$ and estimate when it will become of order 1. We find that for instantons of size
$\rho = 2 a$ ($3a$) the level-crossing will take place for extremely large values of 
the lattice volume, $N_s \sim 500$ ($N_s \sim 3000$).

The situation is much worse for periodic boundary conditions since in that case the 
gap in the spectrum goes to zero with decreasing $\rho /L$.
The results are presented in figure~\ref{fig:iper_lowest}, where we display 
the two lowest eigenvalues of the filtering operator.
The AFM eigenvalue, displayed on the left plot,  matches the twisted 
result, Eq. (\ref{eq:l0_tw}), 
for small $\rho /L$. It becomes however strongly affected
by finite size effects for $\rho \gtrsim L/5$, as a consequence of the absence of exact $Q=1$ solutions
in the periodic case. The first excited eigenvalue, displayed on the right plot, goes to zero as $\rho^2/L^4$.
The fit in the figure corresponds to:
\be
\lambda_1 = \alpha \Big (\frac{\pi \rho}{ L^2} \Big )^2 \quad , 
\label{eq:l1_per}
\ee
with $\alpha = 1.2(1)$. Combining this fit with the one describing the lattice artifacts  Eq. (\ref{eq:l0_tw}),
we can again estimate when the ratio between the two eigenvalues becomes of order 1. In this case it happens
for rather small values of $N_s$: for $\rho = 2 a$ ($3a$) the level-crossing 
will take place at $N_s \sim 20$ ($N_s \sim 35$).

\subsubsection{Two instanton configurations}
\label{s.q2}

In this subsection we will study other classical solutions. 
As in the previous case, the filtering method is bound to work 
in the continuum, so that our goal is mostly that of exploring 
possible problems associated to the discretization. This is a necessary 
step previous to any application to Monte Carlo generated
configurations. Our study has focused in one situation which might
potentially cause problems. This occurs whenever the gauge field
configuration contains widely separated structures. It is to be
expected that in this case the gap narrows down, since every isolated
structure should give rise to a quasi-zero-mode of the $\Opl$ operator.
There is still a unique zero-mode associated to the supersymmetric
zero-mode, but lattice corrections might induce unwanted mixing,
whose effects we want to explore.  

For definiteness we have concentrated  our study in  the case of two 
instantons (I-I) at  variable separations. For the analysis we have 
prepared a set of $Q=2$ SU(2) instanton  configurations, by applying
the following engineering techniques.
Typically, we start with a couple of  instantons  on a  lattice of
size $16^4$. The configurations are glued along the time direction,
and subsequently cooled until the total action drops to  a value close
to $S=16 \pi^2$, signalling that the configuration is approximately a
solution of the classical equations of motion. The gluing can be done
at variable separation among the two instantons, and this is not
substantially altered by the cooling process. Finally, we end up with
a collection of $Q=2$ configurations on a $16^3 \times 32$ lattice and
with periodic boundary conditions. Notice that, in contrast to the
$Q=1$ case, in this case there are indeed classical solutions in the 
continuum living in a box without twist.  

\begin{figure}
\centerline{
\psfig{file=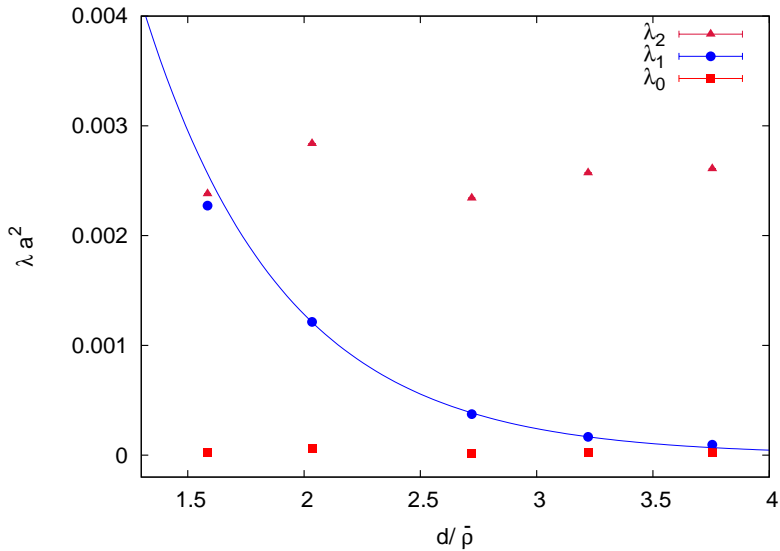,width=0.7\textwidth}
}
\caption{For a set of $Q=2$ instanton configurations, on a $16^3\times 32$ periodic lattice,
we display the three lowest eigenvalues of the $\Opl$ operator
as a function of the inverse I-I overlap: $d/\bar\rho$ (distance over average
size, $\bar \rho$, of both instantons).
The line corresponds to the fit $\lambda_1 a^2= 0.036(5) \exp\{-1.7(1)\,
d/\bar{\rho} \}$. 
}
\label{fig:gapq2}
\end{figure}

We studied the low-lying eigenstates of the $\Opl$ operator for the
afore-mentioned set of configurations. Our results for the spectrum, 
displayed in figure~\ref{fig:gapq2},  agree nicely with our expectations. 
There is a unique zero-mode (up to lattice corrections) for all I-I
separations. The first excited eigenvalue, however,  depends strongly
upon separation $d$, dropping to zero as this separation gets large. 
This dependence can be fitted to  $\lambda_1 a^2= 0.036(5) \exp\{-1.7(1)\, 
d/\bar\rho \}$, where $\bar{\rho}$ is the average size of the two 
instantons. The second excited eigenvalue, on the contrary, 
does not depend significantly upon the separation $d$ and is presumably 
mostly determined by box-size and boundary conditions. 
 
 \begin{figure}
 \centerline{
 \hspace*{3.0cm}\psfig{file=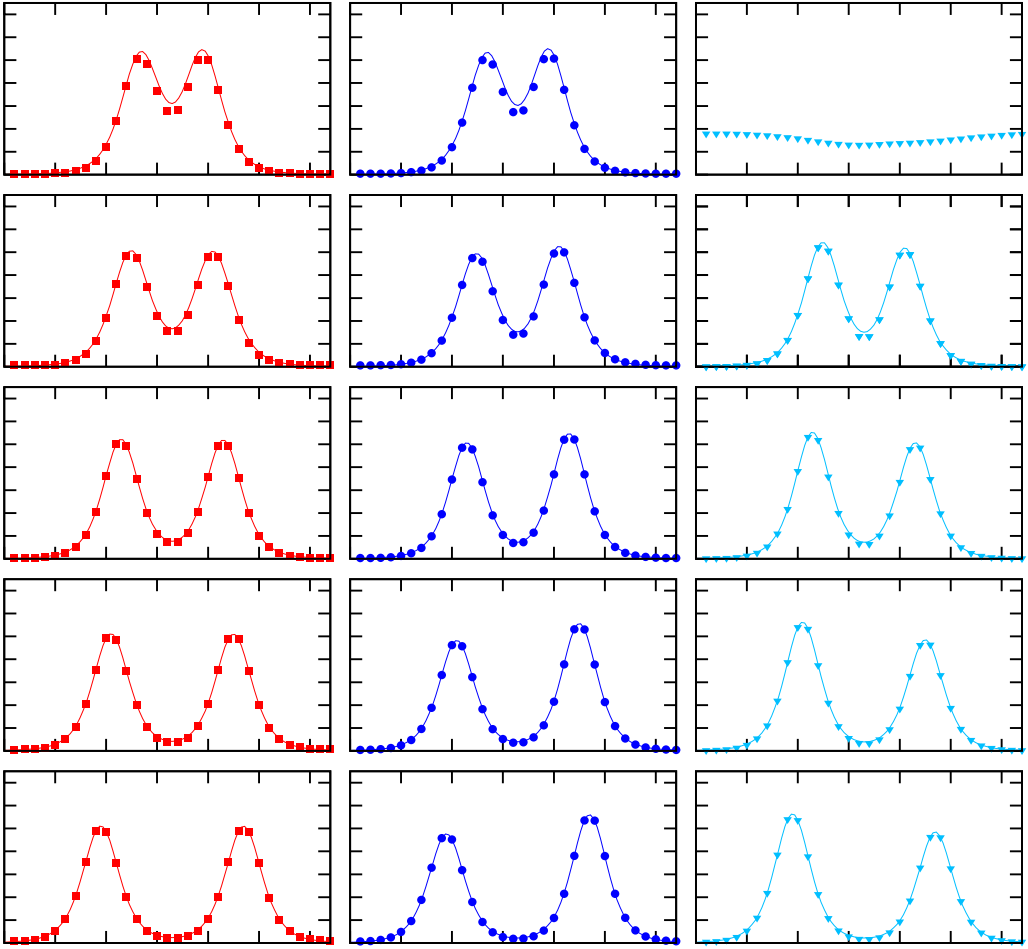,width=1.2\textwidth}
 }
 \caption{For a set of $Q=2$ instanton configurations, on a $16^3 \times 32$ 
periodic lattice, we display from left to right
 the one-dimensional integrated profiles of: the
 self-dual part of the action density; the densities of the lowest and
 first excited
 eigenvectors of the $\Opl$ operator. All the densities are
 normalized in the same way.  The plots from top to bottom correspond
 to I-I separation of  $d = 1.58 \bar \rho$, 2.03$\bar \rho$, 
$2.72 \bar \rho $, $3.22\bar\rho $, and $3.75 \bar\rho $.
 The parameters for the lines in the plot have been obtained from the
fits to the 1-d continuum instanton profiles described in the appendix.
}
 \label{fig:densq2}
\end{figure}
 
Our  interpretation of  the results is substantiated by the structure
of the two lowest eigenvector densities, displayed in
figure~\ref{fig:densq2} for several  values of the separation $d$.
The 1-d profiles represent the integral of the densities over the
spatial directions. For comparison, the spatial integral of the 
self-dual part of the action density is also displayed. Two regimes are
clearly observed. For an I-I distance of $d=1.6\bar{\rho}$, the
situation is similar to the one instanton case: The ground state
density matches the self-dual action density, and the first
excited state corresponds to a nearly constant mode. For larger
separations a level crossing takes place in the excited spectrum and
both densities display the distinctive two instanton profile of the
action density. 

Let us examine these results in the light of our expectations.  In the
limit of infinite separation each instanton ($n=1,2$) has an associated 
supersymmetric quasi-zero-mode $\chi_i^{(n)}= E_i^{(n)}$, given by its
corresponding electric field. As the instantons approach each other, the ground state
of $\Opl$, which  is an  exact zero-mode, is given by the symmetric combination
$\Psi_i^{(0)} \propto E_i^{(1)} + E_i^{(2)}$. The antisymmetric combination should
correspond to the first excited state. Both densities coincide with
the self-dual action density (except for minor corrections in the
region where the instantons overlap). On close  observation, however, we
notice that the density profiles on figure~\ref{fig:densq2} show
slightly different weights for each of the two peaks in both the ground state
and the excited state. This is presumably an effect of the
discretization. Lattice artefacts  induce a mixing of both states,
which can be  estimated  by diagonalizing the
$\Opl$ operator in the subspace spanned by $\Psi^{(0)}$ and $\Psi^{(1)}$. Writing the
perturbed states as:
\bea
\tilde \Psi^{(0)} &=& \cos \theta  \, \Psi^{(0)} - \sin \theta \,
\Psi^{(1)}  \quad,\nonumber \\
\tilde \Psi^{(1)} &=& \ \ \sin \theta  \, \Psi^{(0)} + \cos \theta \,
\Psi^{(1)}  \quad,\nonumber
\eea
one obtains
$$\tan \theta =  \frac{2 \, \delta \Op_{01}}{ \Delta \lambda +
\sqrt{\Delta \lambda^2 +
4 \, (\delta \Op_{01})^2} } \quad,
$$
where $\Delta \lambda$ is the gap between the unperturbed states.
The lattice artefacts enter through the expectation value:
$\delta \Op_{01} = \langle \Psi^{(0)}| \delta \Opl |\Psi^{(1)} \rangle$.
We can estimate the size of $\tan \theta$ by replacing in the
denominator of this
expression the actual gap, extracted from the data in
figure~\ref{fig:gapq2}, and for the
numerator the typical size of the $\Opl$ lowest eigenvalue on a $Q=1$
instanton background. For our I-I configurations we estimate that the
maximal mixing is attained for the largest separation $d=3.75
\bar{\rho}$, being of order  $\tan \theta = 0.05$. A direct
determination based on the relative heights of both peaks gives a value 
of $\tan \theta = 0.04$, in remarkable agreement with our  crude estimate.

Although the possible alteration of the peak heights, for the case of 
sufficiently isolated structures and corresponding small gaps, is a
matter of concern, we want to emphasize that it is not an
insurmountable difficulty, once the origin of the problem is pinned
down. Indeed, our determination of instanton parameters is largely 
insensitive to this normalization. This is clearly shown in
table~\ref{table:q2}, where we collect the  size parameters
$\rho^{(n)}$ of both  instantons  extracted from fitting the action
density and the two  lowest eigenmode  densities. As described
in the appendix, the fitting has been 
done using the shape around the peaks of the 1-d profiles obtained by 
integrating the densities in the spatial directions. The continuous lines in 
figure~\ref{fig:densq2}  correspond
to a superposition of the 1d-instanton profiles given by
Eq.~(\ref{eq:rho1d}) in the appendix, with parameters, including 
normalization, determined from the fits.

\begin{table}
\centerline{
\begin{tabular}{|c|c|c||c|c|c||c||c|}\hline\hline
$\rho_{\AFM}^{(1)}$ & $\rho_{\lambda_1}^{(1)}$ &
$\rho^{(1)}$
&$\rho_{\AFM}^{(2)}$ &$\rho_{\lambda_1}^{(2)}$ &
$\rho^{(2)}$ & $d/\bar{\rho} $ & $\tan \theta$\\ \hline \hline
3.8624 &  -      &3.9201  &3.8045  &  -    &3.8600 &1.58&  - \\ \hline
3.8450 &3.7621   &3.8946  &3.8048  &3.7090 &3.9700 &2.03&0.013\\ \hline
3.6222 &3.6272   &3.6982  &3.6362  &3.6511 &3.7139 &2.72&0.020\\ \hline
3.6641 &3.6531   &3.7315  &3.6717  &3.6630 &3.7498 &3.22&0.036\\ \hline
3.6732 &3.6522   &3.7313  &3.6570  &3.6391 &3.7344 &3.75&0.040\\ \hline
\end{tabular}}
\caption{We give the parameters of the set of $Q=2$ instanton configurations 
analyzed in the paper. All the data correspond to a $16^3\times 32$ lattice with p.b.c.. 
The instanton sizes, in lattice units, are extracted from a fit to the, spatially
integrated,  1-d profiles of the AFM  ($\rho_{\AFM}$), the first excited
state ($\rho_{\lambda_1}$), and the self-dual part of the action density
($\rho$). The separation between the two instantons is given by $d$. 
By $\bar{\rho}$ we denote the average size of the two instantons.
The mixing between the two lowest modes of the $\Op$ operator is given by
$\tan \theta$.
}
\label{table:q2}
\end{table}

To summarize, the filtering procedure succeeds in reproducing the $Q=2$ action density.
We have studied a  distortion in relative normalization of the profiles induced by 
lattice artifacts, which is present when the instanton separation is
large enough.  There might be several ways to cope with it, once we
know its presence and origin. In particular, possible  improvements 
of the $\Opl$ operator, which reduce the size of lattice artefacts,
will also reduce this effect.

\subsection{Smooth non-classical configurations}

Having tested successfully the performance of the filtering method on classical
configurations, we now turn to the analysis of smooth backgrounds that are 
not solutions of the classical equations of motion.
The prototype example is an instanton-anti-instanton (I-A) pair which 
approaches a classical solution only in the limit of infinite I-A separation.
Although in this case we do not expect exact zero-modes of the $\Opm$ operators, 
we will show that one eigenvalue in each chiral sector remains in general
significantly smaller than the others. 
This is so even for configurations in which the instanton and the anti-instanton overlap
considerably and allows to unambiguously select the adjoint filtering modes.
As we will see, the densities of the AFMs correctly reproduce the self-dual and
anti-selfdual parts of the action density.

\begin{figure}
\centerline{
\psfig{file=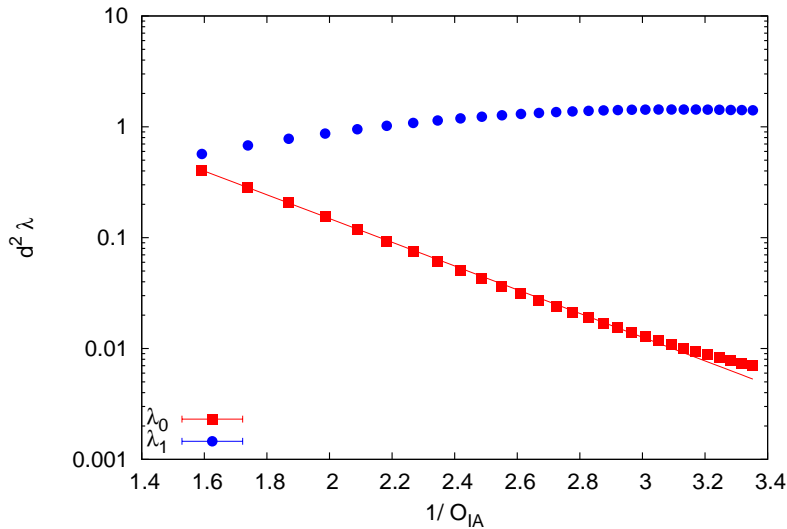,width=0.7\textwidth}
}
\caption{For a set of I-A configurations, on a $12^3\times 24$ periodic lattice, we 
display the two lowest eigenvalues of the $\Opl$ operator 
as a function of the inverse I-A overlap, given by $1/\Ovia= 2d/(\rho_I+\rho_A)$.
The line is a fit to Eq.~(\ref{eq.gapia}).}
\label{fig:specia}
\end{figure}

\begin{figure}
\centerline{
\hspace*{-8cm}
\psfig{file=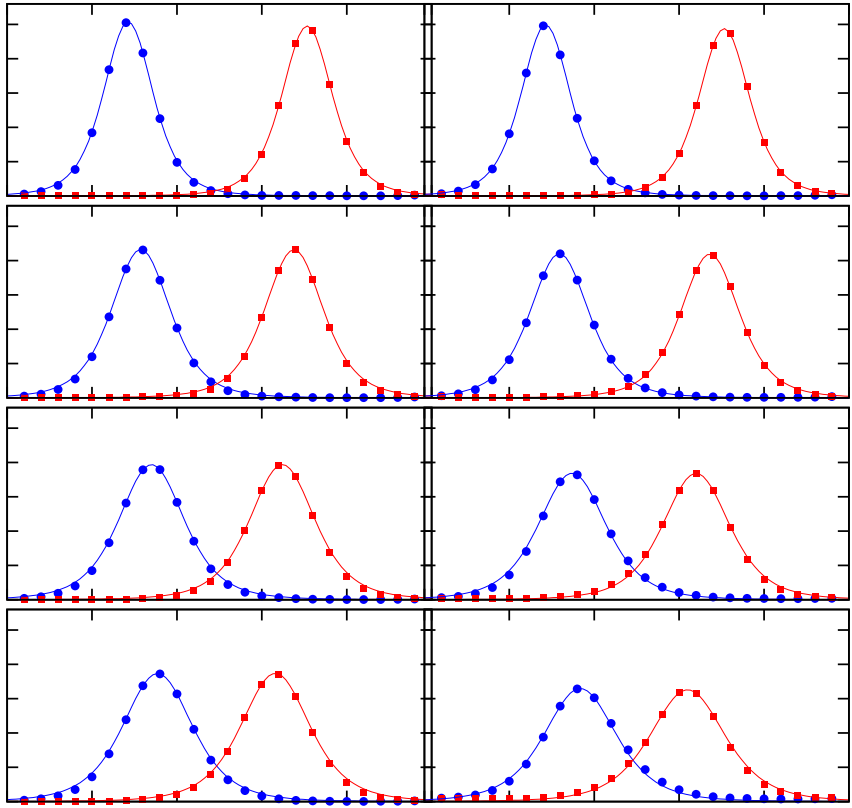,width=0.55\textwidth}
}
\caption{For a set of I-A configurations, on a periodic $12^3\times 24$ lattice, 
we display: Left: the profiles of the self-dual (red) and anti-self-dual (blue) parts of
the action density integrated over the three orthogonal coordinates, and Right:
the same for the densities of the AFM modes on the positive
(red) and negative (blue) chirality sectors. 
All the densities are normalized in the same way.
From top to bottom the plots correspond
respectively to inverse I-A overlap:  $1/\Ovia = 3.35$, 2.42, 1.87, and 1.59.
The lines in the plot correspond to fits to the continuum 1-d instanton profiles. }
\label{fig:densia}
\end{figure}

For the analysis, we have  generated several SU(2) I-A conﬁgurations 
with different I-A separation on a  $12^3 \times 24$ lattice with periodic 
boundary conditions.
The set has been obtained by letting a well separated I-A pair annihilate under 
$\epsilon=0$ cooling. The initial configuration is engineered by gluing along the time
direction an instanton with the anti-instanton obtained from it by time reflection.
In the process of destruction of the pair the action monotonically decreases from
values close to $16\pi^2$ to less than $8\pi^2$, a point after which one can no longer 
identify the I-A lumps in the gauge action density.

Let us present first the analysis of the $\Opm_L$ spectra. 
Figure~\ref{fig:specia} displays the two lowest eigenvalues of the $\Opl$ operator 
as a function of  the inverse I-A overlap: $1/\Ovia= 2d/(\rho_I+\rho_A)$.  
A similar picture is obtained in the negative chirality sector.
The lowest eigenvalue increases with I-A overlap but a clear gap is maintained
as far as $\Ovia<1/1.6$. This allows to clearly identify the AFM.
We will discuss the closing of the gap later on but let us concentrate now on
the AFM densities. They are represented in figure~\ref{fig:densia} for several
characteristic stages in the process of I-A annihilation 
corresponding to $1/\Ovia = 3.35$, 2.42, 1.87, and 1.59.
We present on the left the 1-dimensional profiles of the self-dual 
and anti-selfdual parts of the action density compared with the AFM densities 
displayed on the right.  The agreement is in all 
cases very good, even for the lowest plot that corresponds to the maximal I-A overlap
in figure~\ref{fig:specia}, above which level crossing takes place.
We extract the (anti) instanton parameters by fitting 
these 1-d profiles following the procedure described in the appendix. 
The results are presented in table~\ref{table.ia}, and the corresponding
fits represented by the continuum lines on the plot. 
Even for the extreme case with largest overlap, the difference between 
the size parameters extracted for the AFM modes and the gauge densities
amounts only to 10$\%$, decreasing down to less than 1$\%$ for $\Ovia= 1/3.35$.

\begin{table}
\centerline{
\begin{tabular}{|c|c|c|c|c||c|}\hline\hline
$S/8\pi^2$&$\rho_{\AFM}^I$ & $\rho_{\AFM}^A$ &
$\rho^I$ &  $\rho^A$ & $1/\Ovia$ \\ \hline \hline
1.965&3.2116   &3.1240     & 3.1941   & 3.1045   & 3.35\\ \hline
1.906&3.8008   &3.7744     & 3.7489   & 3.7223   & 2.42\\ \hline
1.787&4.3250   &4.3028     & 4.1402   & 4.1105   & 1.87\\ \hline
1.659&4.8835   &4.7766     & 4.3804   & 4.3500   & 1.59\\ \hline
\end{tabular}}
\caption{We give the parameters of the set of instanton-anti-instanton configurations
analyzed in the paper. All the data correspond to a $12^3\times 24$ lattice with
p.b.c..
The instanton sizes, in lattice units, are computed from the 1-d density profiles of the
positive ($\rho_{\AFM}^{I}$) and negative  ($\rho_{\AFM}^{A}$) chirality AFM modes, and
the self-dual ($\rho^{I}$) and anti-self-dual ($\rho^{A}$) parts of the action density.
The I-A overlap is given by $\Ovia$.
}
\label{table.ia}
\end{table}

One final remark is that, unfortunately, the level-crossing
is controlled not only by the growth of the lowest eigenvalue $\lambda_0$ with
the I-A overlap, but also by the finite volume dependence of the first excited
eigenvalue $\lambda_1$. We can estimate in general when it takes place.
The lowest eigenvalue can be fitted to
\be
d^2 \lambda_0 = 21(1)  \exp\Big(-\frac{ 2.5(1)}{ \Ovia}\Big)\quad ,
\label{eq.gapia}
\ee
where the fit corresponds to the continuum line in figure~\ref{fig:specia}.
The first excited eigenvalue is expected to behave as $1/L^2$ for twisted,  
and as $ \rho^2/L^4$ for periodic boundary conditions
(see respectively Eqs.~(\ref{eq:l1_tw}) and ~(\ref{eq:l1_per})). 
By requiring  $\lambda_0 < \lambda_1$ we obtain:
\bea
L^2 &\lesssim&  \frac{\pi d \rho }{ 4}  \, \exp \Big( \frac{5 d }{4 \rho }\Big)
  \, , \quad  {\rm for\quad  p.b.c.} \quad ,\\
L^2 &\lesssim&  \frac{\pi^2 d^2}{20} \,  \exp \Big( \frac{ 5 d }{ 2 \rho }\Big)
\, , \quad  {\rm for\quad  t.b.c.} \quad .
\eea
For example, the identification of a pair with $d= 1.6 \rho$ requires
$L \lesssim  2 d $ for  periodic boundary conditions, and
$L \lesssim  5 d $ for the twisted case. Once again the use of twisted boundary
conditions pushes away the region in which level crossing takes place.

\section{Testing the method with noisy configurations}
\label{s.heat}
We now come to the crucial test on the filtering method. Our ultimate
goal is to apply the method to Monte-Carlo ensembles and extract from
it a clear picture of its underlying long-range topological content. 
The only way in which one could trust the results is if one has
previously tested the method in a controlled situation. It is not just a 
question of recovering the long-range information stored in the
configuration,  but also of making  sure that the filtering procedure does 
not introduce important distortions in the main parameters describing
the underlying topological structure. 

Our proposed strategy is the following. Starting from a smooth
configuration, we begin to apply Monte Carlo updating steps to it.
We limit ourselves to large values of $\beta$ and a small number of Monte-Carlo
heat-bath sweeps to guarantee that the initial configuration is not destroyed
in the process, and that only statistical noise is added to it. 
In this and the following section we will test the ability of our filtering method
to eliminate the stochastic noise present in heated configurations,
and make use of the information provided to monitor the thermalization
process itself.

\begin{figure}
\centerline{
\psfig{file=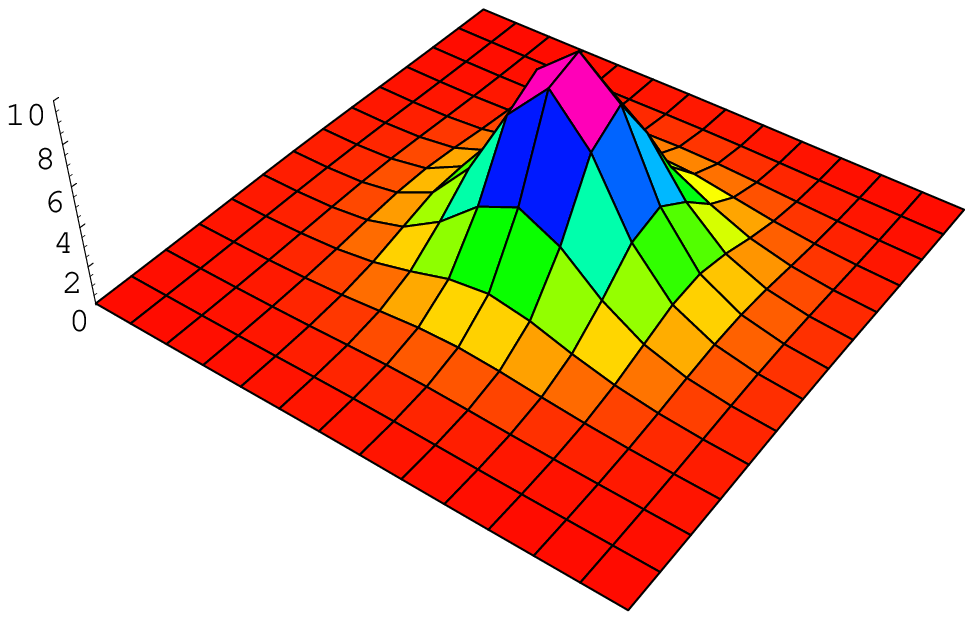,angle=0,width=0.33\textwidth}
\psfig{file=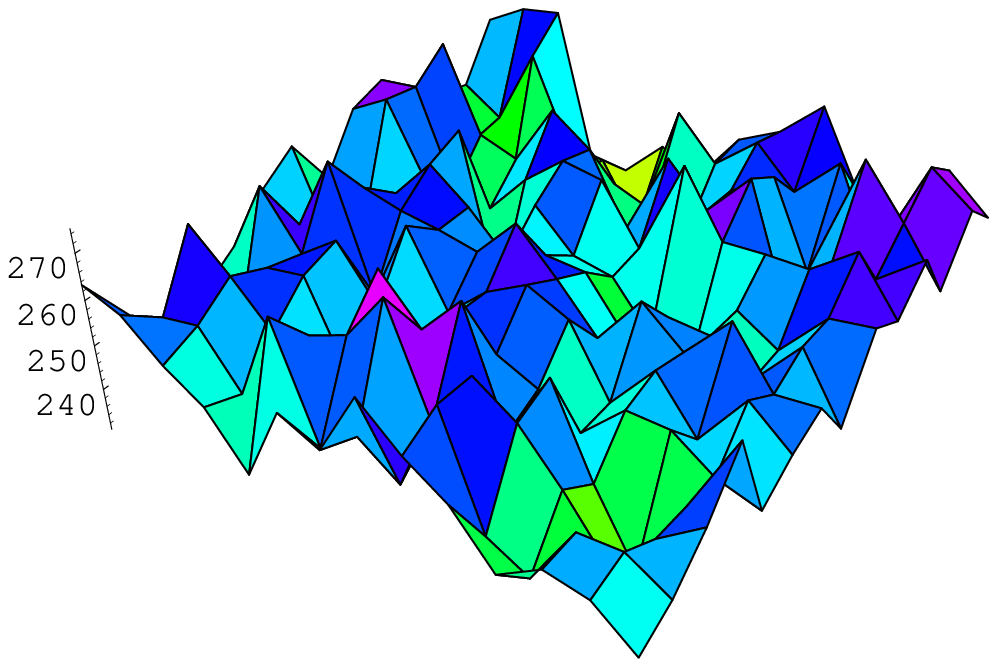,angle=0,width=0.33\textwidth}
\psfig{file=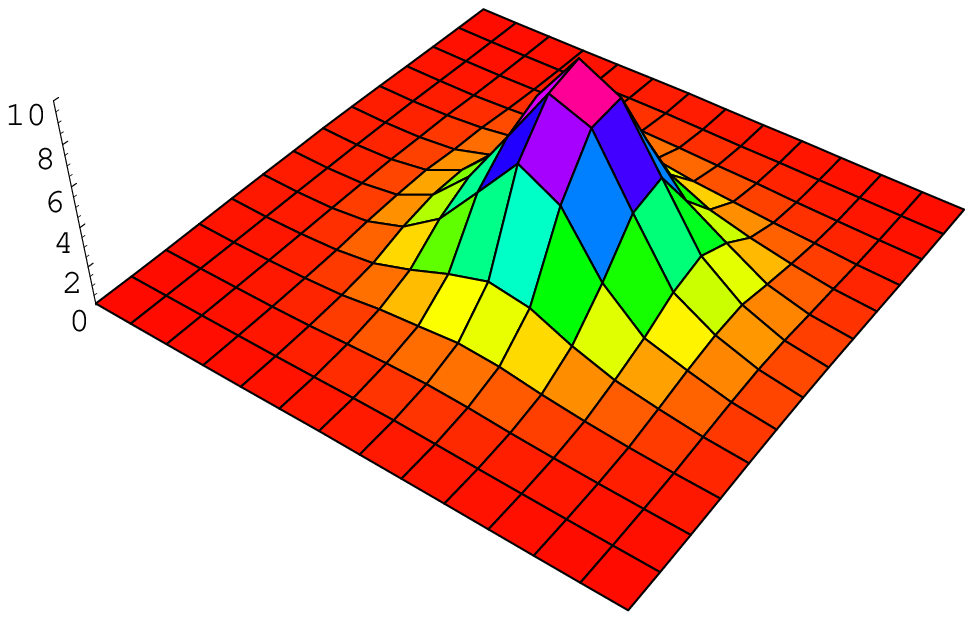,angle=0,width=0.33\textwidth}}
\centerline{
\psfig{file=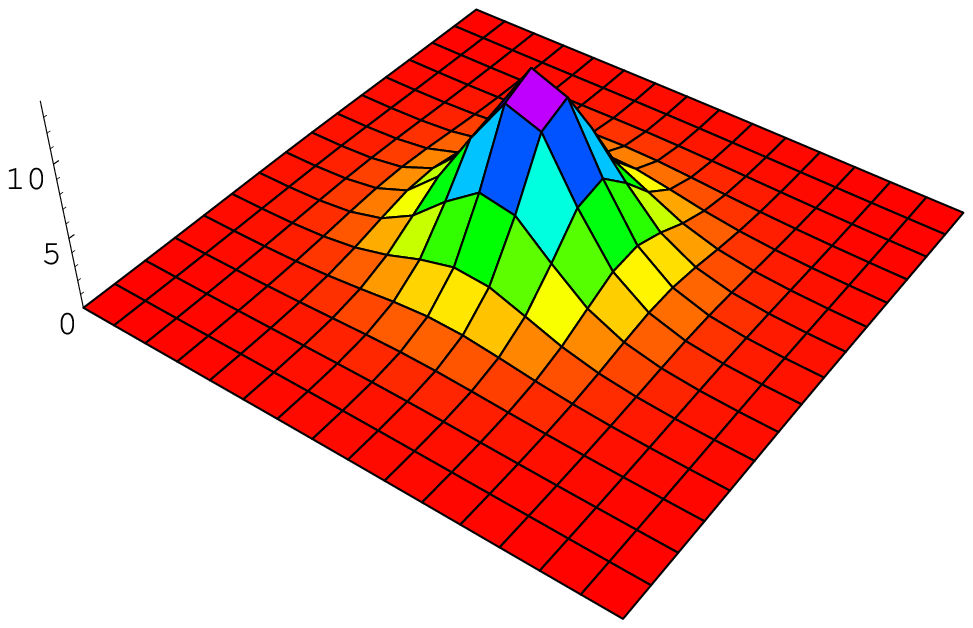,angle=0,width=0.33\textwidth}
\psfig{file=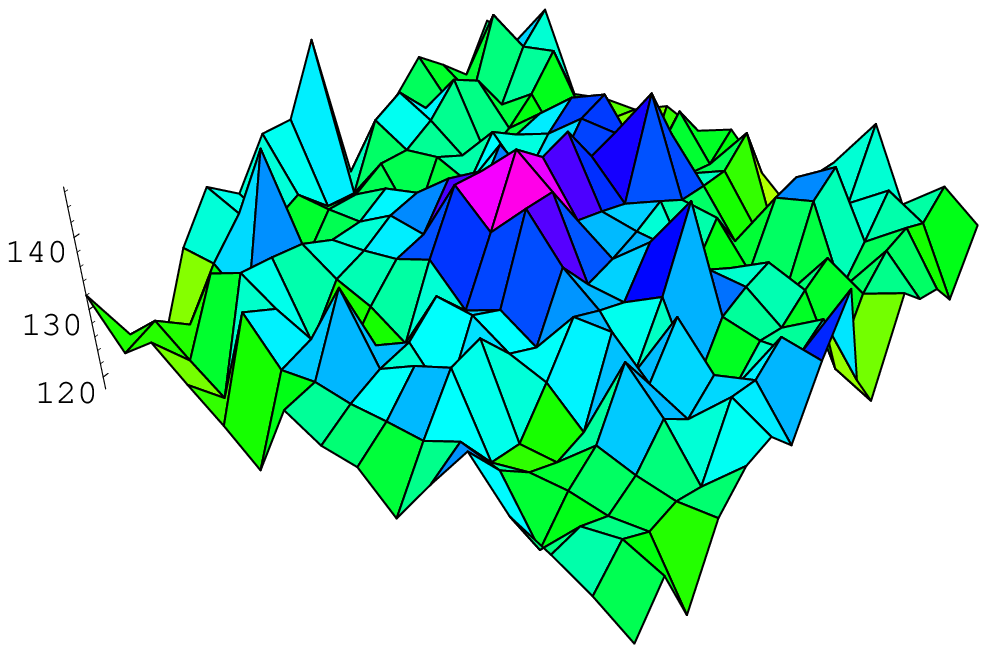,angle=0,width=0.33\textwidth}
\psfig{file=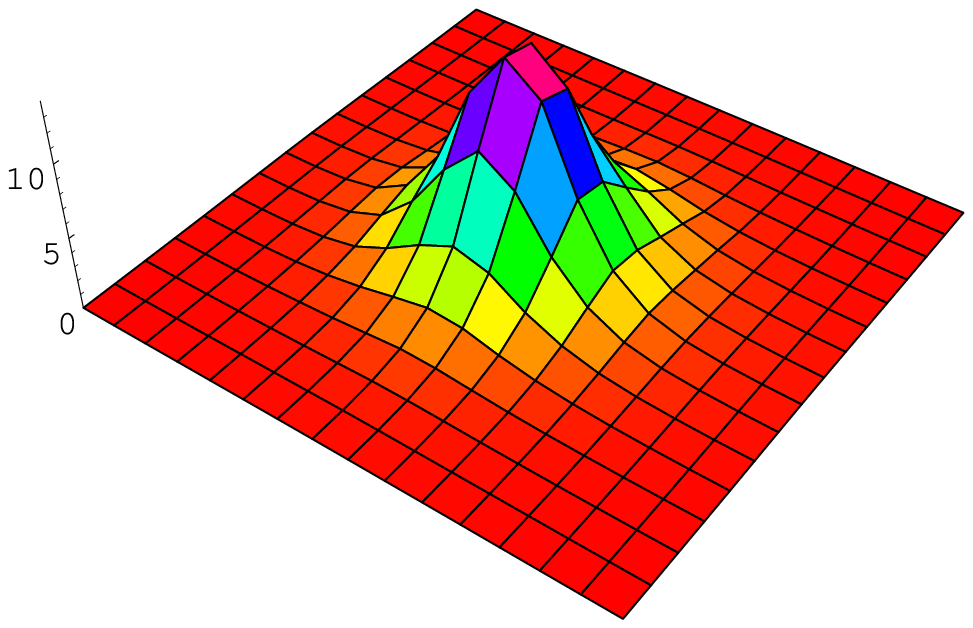,angle=0,width=0.33\textwidth}}
\caption{We display from left to right: the self-dual part of the action density
(in units of  $8\pi^2$) of a smooth $Q=1$ twisted instanton
configuration, the same quantity after applying $N_h$ heat bath sweeps to it,
and the density of the AFM on the heated configuration.
All quantities are integrated in two dimensions.
The top and bottom plots correspond respectively to:
$N_s=14$,  $N_h=10$, and $\beta=3$;  $N_s=16$,
$N_h=60$, and $\beta=8$.
}
\label{fig:dens_h}
\end{figure}

In practise, we have concentrated most of our work in pure SU(2)
gauge theories. Our initial configurations are in all cases  $Q=1$
instantons of various sizes obtained with periodic or  twisted boundary
conditions. If the size parameter  of the instanton is small compared to 
the size of the four-dimensional box no significant change of the
instanton shape is expected with respect to the infinite volume
continuum instanton. 

We have applied $N_h$  heat-bath sweep updates to these smooth
configurations. Short wavelength stochastic noise comes in very
fast  and short range observables (as the average plaquette
value) quickly approach values close to those corresponding to
thermalized ensembles. The size of the noise is of order
$1/\sqrt{\beta}$, but even for relatively large values of $\beta$ 
it dominates the value of the action density, and the trace of the
initial instanton appears to be washed out. This can be clearly 
observed  in figure~\ref{fig:dens_h}. The left pictures display the
self-dual part of the action density of a smooth instanton integrated
in two directions.  The center figures display the same quantity after applying $N_h$
heat-bath sweeps to it, at two different values of $\beta$ ($\beta=3$ with $N_h=10$ for 
the top figure, and $\beta=8$ with $N_h=60$ for the bottom one). Notice the tremendous
change in scale which follows from the size of stochastic noise, even
at these relatively large values of $\beta$. Nevertheless, the
original instanton is still there.  The right figures display the image 
obtained after applying our filtering method to the configurations 
depicted in the center figures. Remarkably, the shape looks rather similar
to the one of the original instanton, and certainly not noisier. 

Although the results look spectacular we are not satisfied by just
obtaining a clean smooth picture of the underlying instanton. We aim
at having full control of the possible distortions introduced in the
parameters (position and size) of the initial instanton. This will be
analysed in the next section, where we will describe the ability of
the filtering method to extract the stochastic motion performed by
these moduli parameters under heating. 

On the basis of the perturbative analysis of section~\ref{s.main} we knew
that the filtering method should act appropriately for configurations 
which are deformations of solutions of the classical equations of motion. 
However, why should  the method still work so well in these cases  in which
the size of the noise is rather large? In the remaining of this
section we will provide a partial answer to this point.  

\begin{figure}
\centerline{
\psfig{file=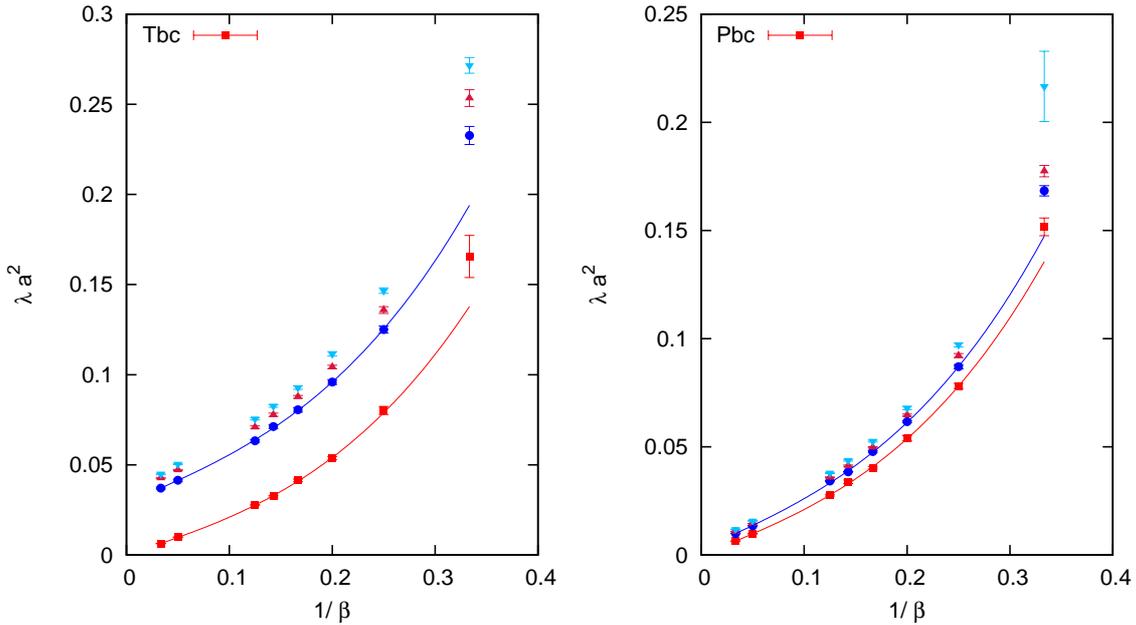,width=1.0\textwidth}}
\caption{We display, as a function of $1/\beta$, the four  lowest
eigenvalues
of the $\Opl$ operator
for the ensemble of heated instantons described in the text. Left and
right  plots correspond  to twisted and periodic boundary
conditions respectively.  Each data point
represents the average of the corresponding eigenvalue over a sample
of ten configurations obtained by applying $N_h$ heat-bath sweeps to a smooth
$Q=1$ instanton. The errors are the dispersion over the sample set.
The continuum lines are fits of the data to a third degree polynomial
in $1/\beta$ - see Eqs.~(\ref{eq.heattbc})-(\ref{eq.heatpbc}).
}
\label{fig:spec_h}
\end{figure}

For that purpose, we studied the  effect of heating on the
low-lying spectrum of the operator $\Opl$. The initial smooth
configuration was taken to be an SU(2) instanton of size $\rho=3.5 a$
on a lattice of length  $N_s=14$, employing  both periodic and  twisted boundary
conditions for the reasons explained previously. 
Starting from these smooth configurations we performed $N_h=10$ heat-bath 
sweeps with the  Wilson action at various values of
$\beta$, ranging from $\beta=3$ to $\beta=30$. For each initial
configuration and value of $\beta$ we repeated the procedure 10 times
with independent random seeds, and averaged the  values obtained for
the four lowest  eigenvalues of $\Opl$ over them. The results  are displayed in
figure~\ref{fig:spec_h}. The continuous line gives the result of a fit 
to the two smallest eigenvalues up to $\beta> 3.3$ for each boundary condition. For
twisted boundary conditions we get:
\bea
\lambda_0^{tbc} &=& \frac{1}{ a^2} \Big [0.190(2) \frac{1 }{ \beta}
+ 2.02(7) \frac{1 }{ \beta^3} \Big ]\quad , 
\label{eq.heattbc}\\
\lambda_1^{tbc} &=& \frac {1 }{a^2} \Big [ 0.57(1) \, \frac{\pi^2 }{ N_s^2} 
+ 0.24(1) \frac{1 }{ \beta} + 2.3(3) \frac{1  }{ \beta^3} \Big ]\quad ,
\eea
and for periodic boundary conditions we obtain:
\bea
 \lambda_0^{pbc} &=& \frac {1 }{a^2} \Big [ 0.191(2) \frac{1 }{ \beta}
+ 1.94(5) \frac{1 }{\beta^3}\Big ]\quad , \\
 \lambda_1^{pbc} &=&  \frac {1 }{a^2} \Big [ 0.9(2) \, \frac{\pi^2 \rho^2 }{L^2  N_s^2} 
+ 0.213(9)\frac{1 }{ \beta} + 1.98(25) \frac{1}{ \beta^3}\Big ]\quad .
\label{eq.heatpbc}
\eea
Notice that as $\beta$ goes to infinity the results approach those obtained for
smooth instantons in the previous sections.  There is a 
zero-mode (up to lattice cut-off effects) for both boundary conditions, 
and a gap which is volume dependent and strongly dependent on boundary
conditions. For large values of $\beta$, the lowest eigenvalue is
proportional to $1/\beta$ as predicted by perturbation theory. The leading correction to 
the higher eigenvalues seems to behave in the same way. Perturbation theory
allows that deformations lead to a modification of the excited  eigenvalues
proportional to $1/\sqrt{\beta}$. Nevertheless, since the noise
introduced by the heat-bath method is stochastic and has both signs,
we find it quite possible that in this case the correction to the
eigenvalue is also of order $1/\beta$ as our data seem to indicate. 
Furthermore, it is remarkable that a fit to the $\beta$ dependence,
depicted as a continuum line in figure~\ref{fig:spec_h}, shows that the
coefficient of the $1/\beta$ term is very much compatible for the first and
second eigenvalue and both boundary conditions. The same applies for
the coefficient of the $1/\beta^3$ term. There is no particular reason 
why there should be no $1/\beta^2$ term as well, but adding it to the
fit gives a coefficient which is compatible with zero, and a larger
value for the $\chi^2$ per degree of freedom.

Some final comments are in order. The lowest eigenvalue of the $\Opl$
operator
depends mildly on  the value of $N_h$. Typically, it  starts to rise
for the initial heat-bath steps, but quickly saturates to a fixed
value. The value also depends slightly on the choice of the heat-bath
action. Taking the improved $\epsilon=0$ action instead of the Wilson 
action only affects the coefficients by 10\%.

The most remarkable feature shown by figure~\ref{fig:spec_h}, is that
the gap remains approximately constant for all values of $\beta$. This 
is the reason why our method works well in all that range. Had level
crossing occurred, our filtered mode would have lost the trace of the 
underlying instanton, and failed dramatically. In the following section 
we will go one step further and investigate how the heat-bath  process
and the filtering affect the instanton parameters.

\section{Analysing Monte Carlo dynamics}
\label{s.dynamics}

In this section we will present the results of a detailed study of the
effect of heat-bath updates and subsequent filtering upon the
collective parameters describing an instanton: its position and size
parameter $\rho$. The philosophy is similar to that presented in the
previous section but taking spatial care in minimising systematic
errors and increasing the statistical significance of the results. 

Our initial smooth instanton configuration was  generated by
$\epsilon$-cooling~\cite{Garcia Perez:1993ki} with $\epsilon=0$ in order to reduce lattice
artifacts. Different instantons sizes have been studied, but most of our data has
been obtained for $\rho=3.5 a$. This is a safe intermediate value,
far from the $\rho=2 a$, in which the instanton soon evaporates and the 
lattice action is strongly dependent of the $\rho$ value, and the larger
values where finite volume starts affecting the shape and  stability. 
The lattice size was taken as $N_s=16$ and we  used twisted boundary 
conditions ($\vec{m}=\vec 0$, $\vec{k}=(1,1,1)$) for the reasons explained before. 

The Monte Carlo updating procedure consisted on  a heat-bath
algorithm based on the classically improved  $\epsilon=0$ lattice
action described previously. Two values of  $\beta$, $16$ and $8$,
were used in our analysis. These relatively large values of $\beta$ 
ensure that the heat-bath updating mostly adds stochastic noise 
to the configuration, but does not destroy it or create new structures. 
Having two values of $\beta$ allows to quantify how much of these
fluctuations are gaussian, allowing us   to extrapolate the results
for lower values of $\beta$. Finally, we have studied the evolution of
the sample for different number of heating steps $N_h$, by applying 
the filtering procedure to all the noisy configurations at
specific values of $N_h$ and extracting 
from the AFM density the value of the instanton parameters, using the methodology
described in the appendix. 

How should the distribution of instanton parameters look like for the 
heated sample? It is quite clear that, in addition to generating gaussian
fluctuations, the heat-bath dynamics should also generate modifications of
the moduli parameters characteristic of the  underlying smooth configuration.
In particular, for instantons, both the location and size parameter $\rho$ 
should behave stochastically as a result of the updating procedure. Up to lattice 
artifacts the position should describe a random walk with no drift term. 
On the other hand, the quantum fluctuations should not be flat in $\rho$ space, because
it is precisely the quantum fluctuations introduced by the updating
procedure what leads to the size dependence of the free energy
computed by `t~Hooft~\cite{'tHooft:1976fv}. Although, the updating procedure is discrete, we
expect that for sufficiently large values of  $\beta$ and of the
number of  updating steps the evolution could be well modelled by a Langevin
equation. The equation for the displacement parameter being purely a 
Wiener process, with a probability distribution satisfying the
heat equation. More precisely, if we started with a gaussian distribution,
it will remain gaussian at all times, with constant mean value and 
a dispersion behaving as 
\begin{equation}
\sigma^2(t)=\langle (x_i(t)-x_i(0))^2 \rangle = \sigma^2(0) + t K_x\quad .
\label{eq.xdisp}
\end{equation}
The slope parameter $K_x$ measures the velocity of the diffusion process
and is a characteristic of the updating procedure. On general grounds
we know, however, that it is proportional to the square of the size of the 
noise. Thus, for our particular case it should go like $g^2\propto
1/\beta$.

\begin{figure}
\centerline{
\psfig{file=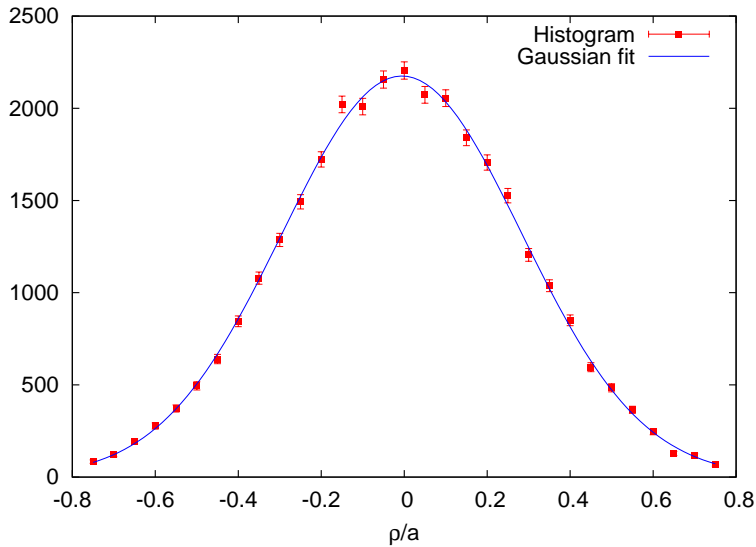,width=0.7\textwidth}}
  \caption{Histogram of position displacement at $N_h=40$ and
  $\beta=16$. The solid line is the fit to a gaussian.}
\label{histpos}
\end{figure}

\begin{figure}
\centerline{ 
\psfig{file=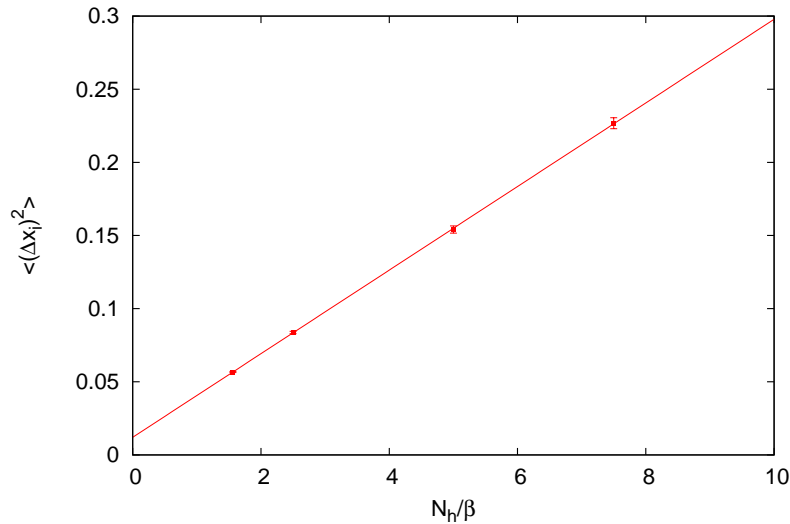,width=0.7\textwidth}}
  \caption{Dispersion in displacement as a function of
    $\frac{N_h}{\beta}$ for $\beta=16$. The solid line is a linear fit.}
    \label{pos_disp}
\end{figure}

Let us see how our data behaves with respect to these expectations. For
that purpose,  starting from our smooth  instanton of size
$\rho/a=3.5$, we  generated 7885 independent trajectories by applying a sequence of
up to 120  heat-bath steps with $\beta=16$. We stored the
configurations after $N_h=$25, 40, 80 and 120 heat-bath sweeps. To the four
sets of 7885 configurations we applied our filtering procedure. In all
cases the filtered image showed a clear peak close to the position of
the initial instanton similar to that displayed in  figure~\ref{fig:dens_h}. 
For each filtered image we extracted the position and size of the corresponding
instanton following the procedure specified in the appendix. Finally,
we performed an statistical analysis of the results. In particular, we
computed the displacement of the center of the instanton obtained from the AFM
relative to the position of the initial instanton.
The data agrees nicely with the expectations of a gaussian
distribution. For example, figure~\ref{histpos} shows the statistical
distribution for $N_h=40$ of the displacement putting the 4 components of the 
displacement in the same plot to increase the statistics.  The data
are compared with a gaussian distribution fitted to the data. The
$\chi^2$ per degree of freedom is 1.15. For the remaining values 
of $N_h$ similar results were obtained with $\chi^2$-pdof of order 1. 
Finally, the mean value and dispersion of the distribution were obtained
either directly or by a fit to a gaussian distribution, with compatible results.
The mean value obtained was always smaller than 0.01 lattice spacings
with no particular $N_h$ dependence. On the contrary, the dispersion
was observed to grow linearly with the number of heat-bath steps. 
Figure~\ref{pos_disp} shows the values of the dispersion as a function of $N_h/\beta$. The
solid line is the result of a fit to a straight line having $\chi^2$
per degree of freedom smaller than 1. The fit is
given by: 
\begin{equation}
\langle (\Delta x_i)^2\rangle = A_x + B_x \frac{N_h}{\beta}\quad ,
\end{equation}
with $A_x= 0.0120(3)$, and $B_x= 0.0286(2)$. The relation of $B_x$ with the 
slope parameter $K_x$ entering Eq.~(\ref{eq.xdisp}) is given by $N_h B_x = t \beta K_x$. 
All results are given in units of the lattice spacing.  

 To test the $\beta$ dependence we repeated the procedure for $\beta=8$,
 $N_h=20$, $40$ and $60$ heat-bath sweeps, and 3940 trajectories.
 The results are qualitatively the same as described for $\beta=16$.
 The  dispersion is fitted to $0.022(3)+0.032(2)  \frac{N_h}{\beta}$.
 Notice that the slope is compatible with the one obtained at
 $\beta=16$, in agreement with the hypothesis of gaussian fluctuations. One feature that
 differs from our Langevin model is the fact that the intercept is
 non-zero.  It is not surprising that the
 continuous Langevin description does not apply during the first few 
 steps. Indeed, the agreement is expected only in the limit $\beta
 \longrightarrow \infty$ with $N_h/\beta$ fixed. This seems to affect 
 mostly the intercept. Apparently the intercept  behaves as 
 $0.19/\beta$. In any case, this small change is the only one which
 could reasonably be attributed to the distortions associated to the 
 filtering method itself.
 
 Now we proceed to the analysis of the $\rho$ evolution. In this case
 we expect a drift due to the $\rho$-dependence of the free energy.
 According to the calculation of Ref.~\cite{'tHooft:1976fv} the
 probability distribution of the size parameter should behave as
 $$ \frac{1}{\rho^5} \rho^{22/3} = \rho^{7/3}\quad . $$
 The first factor comes from the scale invariant distribution in
 collective coordinates, while the second one follows from the
 exponential of the action with a renormalized coupling constant computed at 
 the scale $\rho$. Indeed, the value 22/3 follows from the leading-order 
 beta function. The free energy is then given by minus the logarithm
 of this term. Thus, the Fokker-Planck equation which describes the
 stochastic evolution of the distribution in $\rho$ is given by 
 \begin{equation}
 \partial_0P(\rho,t)= \frac{K_\rho}{2}\, \frac{\partial}{\partial \rho}\left( \frac{7}{3
 \rho} + \frac{\partial}{\partial \rho}\right) P(\rho,t) \quad .
 \end{equation}
In this case, even if one starts with a gaussian distribution in
$\rho$ at time equal zero, it will cease to be gaussian at longer
times. Nevertheless, numerical integration of this equation reveals
very small deviations from gaussianity during the range of times corresponding to 
our heat-bath updates. The drift term shows up in the prediction that
the mean value $\langle \rho\rangle$ becomes an increasing function of
time. This function is approximately a straight line in the
appropriate time range. The dispersion, $\sigma^2(\rho)\equiv 
\langle \rho^2\rangle-\langle \rho\rangle^2$, shows also a (very approximate)
linear growth with time, with a slope given by $K_\rho$. One can circumvent even these 
minor deviations by computing $\langle \rho^2\rangle$. According to the Fokker-Planck
equation one can easily deduce that this quantity should grow exactly linearly with
time:
\begin{equation}
\langle \rho^2\rangle(t) = \langle \rho^2\rangle(0) +K_{\rho}' t \quad .
\end{equation}
The corresponding slope should be higher than $K_\rho$ by a factor that 
gives a measure of the beta function:
\begin{equation}
\frac{K_{\rho}'}{K_\rho}= 1+\frac{7}{3}  \quad .
\end{equation}

\begin{figure}
\centerline{
\psfig{file=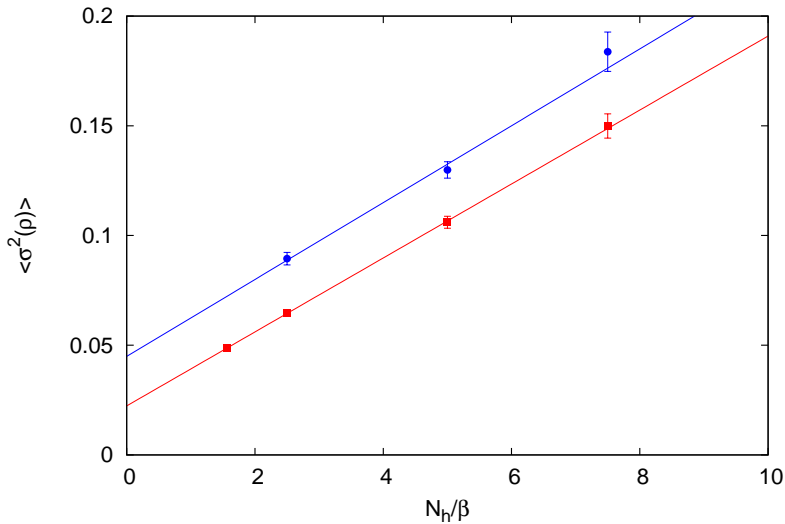,width=0.7\textwidth}}
  \caption{Dispersion in $\rho$  as a function of
      $\frac{N_h}{\beta}$ for $\beta=16$ and $\beta=8$. The solid
      lines are linear
      fits.}
\label{rho_disp}
\end{figure}

\begin{figure}
\centerline{
\psfig{file=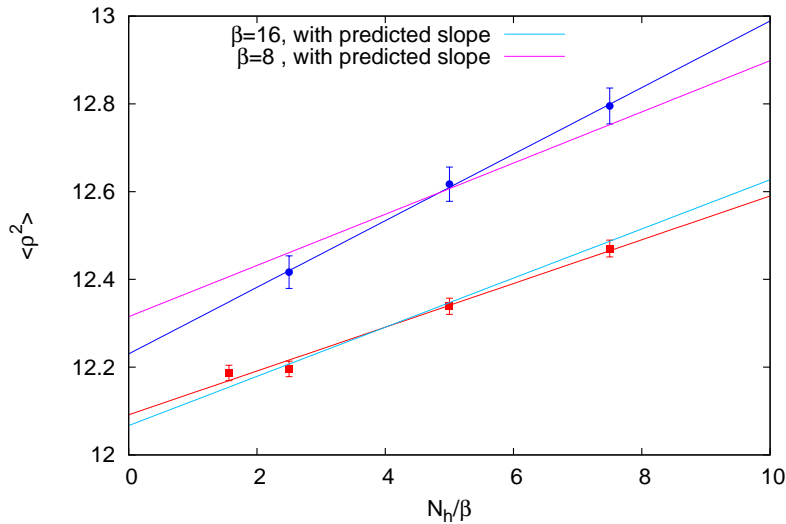,width=0.7\textwidth}}
\caption{The expectation value of $\rho^2$ is displayed as a function
of the number of heating steps over $\beta$. The upper data points
corresponds to $\beta=8$, and the  lower points to  $\beta=16$.  The two
continuum lines for each $\beta$ value correspond to different linear fits,
with either two free parameters or with the slope
fixed to the predicted value: $B'_\rho =10 B _\rho / 3$.}
 \label{rhosq}
\end{figure}

Let us now see how the data behaves with respect to our expectations.
In figure~\ref{rho_disp} we plot the value of  $\sigma^2(\rho)$ for our 
$\beta=16$ and $\beta=8$ data, together with a linear fit to the data. 
The slopes are compatible (0.0168(4) for $\beta=16$ and 0.0175(16) for 
$\beta=8$). Again the intercept scales as $0.36/\beta$. 

The most crucial  aspect is the behaviour of $\langle \rho^2\rangle(t)$
which, according to our expectations, should grow linearly with a slope 
which is 10/3 larger than the one of $\sigma^2(\rho)$. In
figure~\ref{rhosq} a fit to the data is performed showing, for $\beta=16$, 
a slope of $0.050(4)$. This gives a ratio of $8.9(8)/3$ instead of $10/3$. 
This is less than 2 sigma away from the expected value, and gives in any case 
a value incompatible with the one observed for the dispersion $\sigma^2(\rho)$. We can
safely claim that we have obtained numerical evidence for the predicted 
dependence of the free energy with $\rho$. The data at $\beta=8$ show
the same trend, although this time the slope overshoots and the ratio
is rather $13(1)/3$. All the parameters of the linear fit to our
data are collected in table~\ref{table:q8}. The errors quoted include
both statistical (using jack-knife)  and systematic effects coming from 
correlation upon fit-parameters.

We have done other checks at different values of $\rho$ but with
smaller statistics. It is impossible in this case to determine the
drift in the mean value of $\rho$, but given that the dispersions are 
affected by much smaller errors, we can extract  meaningful values 
of $K_x$ and $K_\rho$ from the data. In particular we generated 500 
trajectories at  $\rho/a=2.85$, $3.25$ and $3.784$.  In all cases we
obtain good linear fits for the dispersion in displacement
$\sigma^2(\Delta x)$ and in size $ \sigma^2(\rho)$. The parameters of
the fits are collected in table~\ref{table:q8} and are reasonably
compatible with each other. 
Although, the range in $\rho$ is small
and the errors are large, we can safely conclude that the slopes are 
not strongly dependent on the value of $\rho$.

Our data predicts the dependence of the displacement and $\rho$ as a
function of the number of heating steps. The results can be
extrapolated to lower values of $\beta$. Although, as $\beta$ decreases 
we expect non-linear dependencies in $\beta$ it would be very
surprising that the results change by orders of magnitude. What are
the possible implications of these results? Suppose one has a Monte
Carlo generated configuration with a particular instanton content.
Applying heat-bath  sweeps to this configuration the positions and 
sizes of instantons will change approximately according to the pattern
derived in this paper. Eventually either by displacement or by growth 
the instantons will meet anti-instantons and with a finite probability 
annihilate. The last steps could be very fast, but for large initial
separations most of the time will be employed in the stochastic motion
depicted in this paper. Other processes are possible. Equal sign
instantons might merge into new structures and even percolate as a
result of the growth in $\rho$. In addition, dislocations can be
created which might eventually develop into new instantons and change
the value of the topological charge of the configuration. Finally, one
can have the reverse phenomenon, since small instantons have a finite
probability of decreasing in size and disappearing through the holes
of the lattice. It is clear that only once these processes have acted 
sufficiently can we approach a thermal distribution of the topological
structure. The autocorrelation time for this type of processes are
much larger than those affecting local quantities such as the mean
value of the plaquette. Assuming that $K_x$ does not depend on
$\rho/a$, as our data seems to support, we can conclude that, in order 
to reach changes in displacements  and sizes which are given by appropriate
values $l$ in physical units, we need to perform a number heating steps
given by:
\begin{equation}
N_h = l^2 \beta \frac{1}{a^2(\beta) B} \quad ,
\end{equation}
This grows very fast for large values of $\beta$ due to the presence 
of the $a^2(\beta)$ factor in the denominator. This is a typical
critical slowing down phenomenon. The factor of $1/B$ adds to the
difficulty, since it is of order 30-60 for displacements and sizes 
respectively. If we input the  characteristic values of $l$ and $\beta$ one can 
provide an estimate of the corresponding autocorrelation times. 

\begin{table}
\centerline{
\begin{tabular}{|c|c|c|c|c|c|c|c|}\hline\hline
$\beta$ & $\rho_0/a$  & $\#$ Confs & $N_h$ & $\beta A_x$ & $B_x$ & $\beta A_\rho$ &
$B_\rho$ \\ \hline \hline
16  & 3.5  & 7885  & 25, 40, 80, 120 & 0.192(7) & 0.0286(4) & 0.357(15)& 0.0168(4)\\ \hline
8   & 3.5  & 3940  & 20, 40, 60  & 0.17(2)  & 0.032(2)  & 0.36(5) & 0.0175(16)\\ \hline
16    & 3.25  & 500  & 5, 10, 20, 50, 100 & 0.10(1) & 0.031(1) & 0.20(4) & 0.019(2) \\ \hline
16    & 3.78  & 500  & 5, 10, 20, 40 & 0.14(2) & 0.034(2)& 0.28(6) & 0.023(4) \\ \hline
16    & 2.8  &  500& 10, 20, 30, 40  & 0.06(1)  & 0.0314(5) & 0.17(2) & 0.0189(15) \\ \hline
\end{tabular}}
\caption{List of fitted parameters to the dispersion in position
displacement ($ \sigma^2(\Delta x)= A_x+
\frac{B_x N_h}{\beta}$) and $\rho$  values ($ \sigma^2(\rho)=
A_\rho+ \frac{B_\rho N_h}{\beta}$) . The relation with the continuum slope
parameters $K_x$ and $K_\rho$ is given by: $N_h B_x = t \beta K_x$, and
$N_h B_\rho = t \beta K_\rho$.}
\label{table:q8}
\end{table}

\section{Conclusions and Outlook}
\label{s.conclusions}

In this paper we have studied a filtering method proposed in
Ref.~\cite{Gonzalez-Arroyo2006}. We have considered  a simple and precise definition
of the filtered profiles, and implemented  it on the lattice by means
of the overlap Dirac operator in the adjoint representation. We have
tested the method with various classes of initial configurations.
First of all, we analysed  solutions of the classical equations of
motion for which the method should work optimally in the continuum.
The goal here was to determine the effects of lattice artifacts and
the distortions obtained in the extraction of the physical parameters
of the configuration. For the case of instantons these parameters are
the  position and the size parameter $\rho$. Then, we proceeded to study smooth
configurations which are not classical solutions. In particular, we
focused on instanton-anti-instanton pairs at varying
distances. For all types of smooth configurations
the method performs nicely until the configuration is too rough (small
instantons) or very far from a classical configuration (very nearby
instantons and anti-instantons). Widely separated structures might
also give rise to problems, since the gap is reduced. Our  goal in
this study was not only to see that the method works nicely with very
small distortions, but also to identify possible
limitations and problems that can arise when applying the method to a
typical configuration appearing in the QCD vacuum.

Next, we proceeded to add quantum fluctuations by applying  a series of heat-bath
sweeps  to an initial  smooth configuration. To guarantee that the
noise does not destroy the initial topological structure we considered
a lattice inverse coupling $\beta$ larger than $3$ and a small number
of heat-bath steps. The results are quite impressive since the
filtered image looks very smooth and remarkably close to the initial
configuration. The variations in the parameters of the filtered image
with respect to those of the initial configuration cannot be entirely
attributed to distortions induced by  the filtering process. Part of these
variations are due to  genuine stochastic motions of these parameters as a result
of the heating process. Given the good quality of our results, we
embarked in a high statistics analysis of these stochastic motions for
the case of an initial single instanton configuration.
The results obtained are quite precise and follow perfectly the
expectations of a Langevin-equation modelling of this process. In
particular, the time evolution of the dispersions of the instanton
parameters (position and size)   can be used to estimate the characteristic
autocorrelation times of  heat-bath dynamics for producing changes in
the  properties of topological structures present in the configuration.
Furthermore, we see positive evidence of the quantum free energy dependence
on $\rho$ computed many years ago by `t Hooft  in perturbation theory.
Thus, our data  provides a numerical determination  of the first coefficient
of the $\beta$ function of Yang-Mills theory, which is roughly in agreement
with the number extracted from perturbation theory.

Given that the method seems to work very well in all the tests carried
out in this paper, it might indeed become an effective tool to analyse the
topological content of Monte Carlo generated configurations at  typical lattice
spacing scales. Ultimately it could allow to test the validity of all
the proposed scenarios  of the Yang-Mills theory vacuum at finite and zero
temperature. This is, of course, a very ambitious goal which demands a
thorough and careful analysis. Our initial tests look rather promising.

Fortunately, the method has  other applications. In particular,
along the lines presented in  the previous section,  one
can use it to study the dynamics induced by  the Monte Carlo updating procedure
 on characteristic topological structures. There are two aspects
 involved: the shape  of the generated quantum distribution and the
 rate at which this distribution is attained.  With respect to the
 first aspect, the case of the single instanton studied in this paper
 is  the simplest. Perturbation
theory provided long time ago the structure of the induced quantum  weights for small
instantons. However, large instantons, calorons and  fractional instantons are other topological
structures which have a richer parameter space and for which
perturbation theory has only partial answers. For example, for the case
of calorons both masses and positions of the constituent monopoles
determine its parameter space. What is the dependence of the quantum
weights on them? There have been some recent results in perturbation
theory~\cite{Diakonov} addressing these issues. Our method can verify these
results and explore other instances for which perturbative results are
not available.

The second aspect is also very important from the practical viewpoint.
Indeed, different updating procedures might lead to different
autocorrelation times for topological features present in the
configurations. The number of sweeps necessary to generate a
thermalized sample depends crucially on that number. Our method can
then be used to compare different updating procedures and determine
the optimal one. This issue is a rather hot topic which has generated
some debate in the literature, after very large autocorrelation times
were observed for overall topological quantities such as total
charge~\cite{critslow}-\cite{critslow3}.

We want to add some final comments about possible improvements of the
method on the methodological side. Some of them could be
associated with the lattice implementation, by using improved versions
of the Wilson-Dirac operator entering in the construction of the
overlap operator. Other changes can be associated with the definition
of the continuum operator itself $\Opm$. In particular, one could take 
instead operators of the form $-PDK\bar{D}P$. A convenient choice of
the kernel $K$ can preserve all the nice properties of $\Opm$ (positivity,
reality, unique zero-mode for classical solutions, etc) while allowing
to minimize corrections or increase the gap in the spectrum. For the time being
these modifications have proved unnecessary, but this possibility might
be worthy of future exploration.

\acknowledgments
We acknowledge financial support from the MCINN grants FPA2009-08785 and FPA2009-09017, the Comunidad Aut\'onoma
de Madrid under the program  HEPHACOS S2009/ESP-1473, and the European Union under 
Grant Agreement number PITN-GA-2009-238353 (ITN STRONGnet).
The authors participate in the Consolider-Ingenio 2010 CPAN \linebreak
(CSD2007-00042). We acknowledge the use of the IFT clusters for part of our numerical
results.

\appendix
\label{s.appendix}

\section{Instanton size determination}
As explained in the paper, it is important to establish the effect of
the filtering procedure and of the heat-bath algorithm  on the
distribution of instanton parameters: position and size $\rho$. It is
clear that a discretized instanton action density is not identical to the analytic
continuum expression. We expect hence that different lattice determinations  of
instanton position and sizes differ by values of order $a$.
The differences should decrease as $\rho/a$ increases. 
In addition to this, there is also a finite volume correction
which modifies the shape of the instanton as $\rho/L$ approaches 1.
This correction is dependent on the boundary conditions. 

In what follows we will describe the methods we have used in this paper to 
determine the instanton parameters. Although they are affected by lattice corrections,
our concern here is to provide a quantitative comparison between 
two different density configurations (zero-mode or action density, 
filtered or not-filtered, heated or not-heated) employing the same algorithm 
for extracting the instanton parameters in both cases. In this way an important part 
of the lattice corrections disappear when taking the difference.  

We have used two different methods to determine the instanton parameters 
from the lattice densities. The purpose of this appendix is
to explain both of them and compare their results. 

The first method extracts the instanton parameters by comparing the
lattice density field with that of a continuum instanton at infinite
volume, given by:
\be
S(x) = \frac{48 \pi^2 \rho^4 }{  (\rho^2 + (x-X)^2)^4} \quad ,
\label{act_dens}
\ee
where $\rho$ and $X$ denote respectively the instanton size and
position. For that purpose, one fits the lattice density field in the
vicinity of its local maximum to Eq.~(\ref{act_dens}).
One of the nice advantages of this method is that it can be applied to
configurations having several sufficiently dilute instantons. The procedure can
then be applied to all of the local maxima (peaks) and determine all the
parameters of each instanton. Following 
Ref.~\cite{de Forcrand:1997sq}, the method attempts to take into account
the finite volume effects and the interaction with other
instanton-like objects by using an ellipsoid generalization of the instanton 
ansatz given by:
\be
\tilde S(x) = K \Big (1 + \sum_\mu
\frac{(x_\mu-X_\mu)^2}{ \rho_\mu^2}\Big )^{-4} \quad .
\label{eq:rho4d}
\ee
For an exact instanton $K= 48 \pi^2 /  \rho^4$, and 
all the $\rho_\mu$ are equal.
This formula has in total 9 parameters which can be extracted 
by solving for $\tilde S^4(x)$ on the lattice location of the peak 
and its 8 nearest neighbours. From the values of $\rho_\mu$ we extract a
size parameter given by the geometrical average: $\rho^4 \equiv \prod_\mu \rho_\mu$. 
Note that the $\rho$ parameter computed in this way is independent of the 
overall normalization of the density, which is instead encoded in the value
of $K$. 

\begin{figure}
\centerline{
\psfig{file=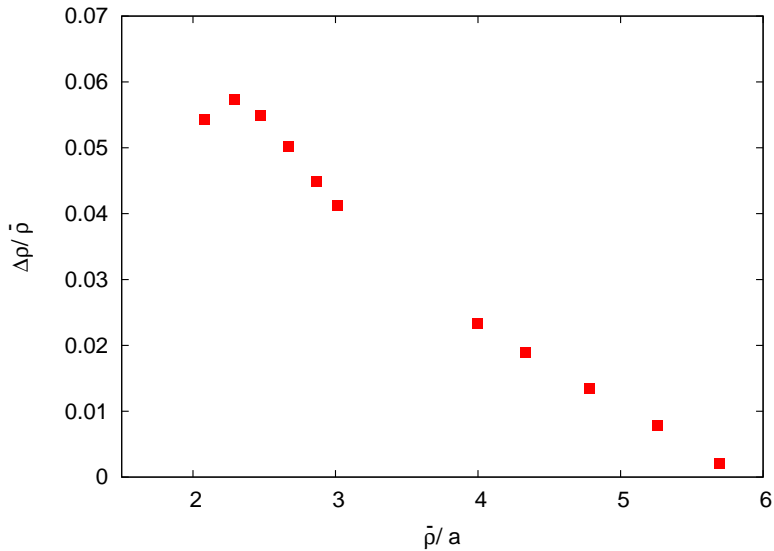,width=0.7\textwidth}}
\caption{
For several $Q=1$ instanton configurations and twisted boundary conditions, 
we display the relative difference between the two size parameters
described in the appendix.  
}
\label{fig:sizes}
\end{figure}

The second method employs the instanton profile in one dimension. This
is obtained by integrating the action density over the three spatial coordinates
giving   
\be
P (x_0) =\int d^3\vec{x} \, {\cal S}(x) 
= \frac{192 \pi}{\rho} \ \int dr
r^2 \Big(1+r^2+\frac{(x_0- X_0)^2 }{ \rho^2}\Big)^{-4} 
=\frac{6 \pi^2}{\rho} \Big(1+\frac{(x_0-X_0)^2}{ \rho^2}\Big)^{-5/2} . \nonumber
\ee
The corresponding lattice quantity $P^L (n_0)$ is obtained by 
summing the lattice action density over three-dimensional volumes. The
parameters $X_0$ and $\rho$ are then determined by identifying the value
of $P^L (n_0)$  at the maximum and its two neighbouring lattice points 
with:
\be
(P^L(n_0))^{-2/5}=\tilde K  \Big (1+\frac{(a n_0-X_0)^2}{ \rho^2}\Big ) \quad .
\label{eq:rho1d}
\ee
As in the previous method, the value of $\tilde K$ is also a function of $\rho$, 
but it is not taken into account for its determination. Thus, this method 
is also insensitive to the normalization of the density.  
The procedure is repeated for all directions and in this way we 
determine the four coordinates of the center, and four different
estimates of the $\rho$ parameter. From them we can extract 
the mean value and the error in the determination of $\rho$.

It is clear that the second procedure is non-local as opposed 
to the first one. This implies that in the case of noisy configurations
the statistical fluctuations are reduced. On the contrary it
fails when applied to an environment in which many instantons
are present. That is the reason why we prefer to employ both methods 
instead of sticking to a single procedure. 
In any case, for instantons that are sufficiently smooth, the error 
in the size determination induced by the discretization remains at 
the few percent level.
This is illustrated in figure~\ref{fig:sizes} where we compare the size
parameters, extracted from the two methods described above, 
for a set of twisted $Q=1$ instanton configurations of varying sizes. 
For instantons larger than $2.5 a$, the difference between the two $\rho$ 
determinations decreases as $\rho/a$ increases and remains in all cases
below $5\%$.


\begin{thebibliography}{}

\bibitem{Callan:1977gz}
  C.~G.~.~Callan, R.~F.~Dashen and D.~J.~Gross,
  ``Toward a theory of the strong interactions,''
  Phys.\ Rev.\  D {\bf 17} (1978) 2717.

\bibitem{'tHooft:1976fv}
  G.~'t Hooft,
  ``Computation of the quantum effects due to a four-dimensional
  pseudoparticle,''
  Phys.\ Rev.\  D {\bf 14} (1976) 3432
  [Erratum-ibid.\  D {\bf 18} (1978) 2199].

\bibitem{Belavin:1976vj}
  A.~A.~Belavin and A.~M.~Polyakov,
  ``Quantum Fluctuations Of Pseudoparticles,''
  Nucl.\ Phys.\  B {\bf 123} (1977) 429.



\bibitem{Shuryak:1981ff}
  E.~V.~Shuryak,
  ``The Role Of Instantons In Quantum Chromodynamics. 1. Physical Vacuum,''
  Nucl.\ Phys.\  B {\bf 203} (1982) 93;
  ``The Role Of Instantons In Quantum Chromodynamics. 2. Hadronic Structure,''
  Nucl.\ Phys.\  B {\bf 203} (1982) 116;
``The Role Of Instantons In Quantum Chromodynamics. 3. Quark - Gluon
  Nucl.\ Phys.\  B {\bf 203} (1982) 140.

\bibitem{Ilgenfritz:1980vj}
  E.~M.~Ilgenfritz and M.~Muller-Preussker,
  ``Statistical Mechanics Of The Interacting Yang-Mills Instanton Gas,''
  Nucl.\ Phys.\  B {\bf 184} (1981) 443.

\bibitem{Diakonov:1983hh}
  D.~Diakonov and V.~Y.~Petrov,
  ``Instanton Based Vacuum From Feynman Variational Principle,''
  Nucl.\ Phys.\  B {\bf 245} (1984) 259.

\bibitem{Schafer:1996wv}
  T.~Schafer, E.~V.~Shuryak,
  ``Instantons in QCD,''
  Rev.\ Mod.\ Phys.\  {\bf 70}, 323-426 (1998) [hep-ph/9610451].

\bibitem{Atiyah:1978ri}
  M.~F.~Atiyah, N.~J.~Hitchin, V.~G.~Drinfeld, Y.~.I.~Manin,
  ``Construction of Instantons,''
  Phys.\ Lett.\  {\bf A65}, 185-187 (1978).

\bibitem{tonyfrac}
 M.~Garc\'{\i}a Perez, A.~Gonz\'alez-Arroyo and P.~Mart\'{\i}nez,
  ``From perturbation theory to confinement: How the string tension is built
  up,''
  Nucl.\ Phys.\ Proc.\ Suppl.\  {\bf 34} (1994) 228
  [arXiv:hep-lat/9312066].
A.~Gonz\'alez-Arroyo and P.~Mart\'{\i}nez,
  ``Investigating Yang-Mills theory and confinement as a function of the
  spatial volume,''
  Nucl.\ Phys.\ B {\bf 459} (1996) 337
  [arXiv:hep-lat/9507001].
A.~Gonz\'alez-Arroyo, P.~Mart\'{\i}nez and A.~Montero,
  ``Gauge invariant structures and confinement,''
  Phys.\ Lett.\ B {\bf 359} (1995) 159
  [arXiv:hep-lat/9507006].
A.~Gonz\'alez-Arroyo and A.~Montero,
  ``Do classical configurations produce Confinement?,''
  Phys.\ Lett.\ B {\bf 387} (1996) 823
  [arXiv:hep-th/9604017].

\bibitem{latticeconf}
  R.~S.~Van de Water,
  ``The CKM matrix and flavor physics from lattice QCD,''
  PoS {\bf LAT2009} (2009) 014
  [arXiv:0911.3127 [hep-lat]].


\bibitem{Albanese:1987ds}
  M.~Albanese {\it et al.}  [APE Collaboration],
  ``Glueball Masses and String Tension in Lattice QCD,''
  Phys.\ Lett.\  B {\bf 192} (1987) 163.
\bibitem{Falcioni:1984ei}
  M.~Falcioni, M.~L.~Paciello, G.~Parisi and B.~Taglienti,
  ``Again On SU(3) Glueball Mass,''
  Nucl.\ Phys.\  B {\bf 251} (1985) 624.
\bibitem{DeGrand:1997gu}
  T.~A.~DeGrand, A.~Hasenfratz and T.~G.~Kovacs,
  ``Topological structure in the SU(2) vacuum,''
  Nucl.\ Phys.\  B {\bf 505} (1997) 417
  [arXiv:hep-lat/9705009].
\bibitem{Hasenfratz:1998qk}
  A.~Hasenfratz and C.~Nieter,
  ``Instanton content of the SU(3) vacuum,''
  Phys.\ Lett.\  B {\bf 439} (1998) 366
  [arXiv:hep-lat/9806026].
\bibitem{Morningstar:2003gk}
  C.~Morningstar and M.~J.~Peardon,
  ``Analytic smearing of SU(3) link variables in lattice QCD,''
  Phys.\ Rev.\  D {\bf 69} (2004) 054501
  [arXiv:hep-lat/0311018].




\bibitem{Teper:1985rb}
  M.~Teper,
  ``Instantons In The Quantized SU(2) Vacuum: A Lattice Monte Carlo
  Investigation,''
  Phys.\ Lett.\  B {\bf 162} (1985) 357.
\bibitem{Berg:1981nw}
  B.~Berg,
  ``Dislocations And Topological Background In The Lattice O(3) Sigma Model,''
  Phys.\ Lett.\  B {\bf 104} (1981) 475.
\bibitem{Hoek:1986nd}
  J.~Hoek, M.~Teper and J.~Waterhouse,
  ``Topological Fluctuations And Susceptibility In SU(3) Lattice Gauge
  Theory,''
  Nucl.\ Phys.\  B {\bf 288} (1987) 589.
\bibitem{Ilgenfritz:1985dz}
  E.~M.~Ilgenfritz, M.~L.~Laursen, G.~Schierholz, M.~Muller-Preussker and H.~Schiller,
  ``First Evidence For The Existence Of Instantons In The Quantized SU(2)
  Lattice Vacuum,''
  Nucl.\ Phys.\  B {\bf 268} (1986) 693.


\bibitem{Garcia Perez:1993ki}
M.~Garc\'{\i}a Perez, A.~Gonz\'alez-Arroyo, J.~R.~Snippe and P.~van Baal,
  ``Instantons from over - improved cooling,''
  Nucl.\ Phys.\  B {\bf 413} (1994) 535
  [arXiv:hep-lat/9309009].

\bibitem{de Forcrand:1997sq}
  P.~de Forcrand, M.~Garc\'{\i}a Perez and I.~O.~Stamatescu,
  ``Topology of the SU(2) vacuum: A lattice study using improved cooling,''
  Nucl.\ Phys.\  B {\bf 499} (1997) 409
  [arXiv:hep-lat/9701012].

\bibitem{GarciaPerez:1998ru}
  M.~Garc\'{\i}a Perez, O.~Philipsen and I.~O.~Stamatescu,
  ``Cooling, physical scales and topology,''
  Nucl.\ Phys.\  B {\bf 551} (1999) 293
  [arXiv:hep-lat/9812006].

\bibitem{Luscher:2010iy}
  M.~Luscher,
  ``Properties and uses of the Wilson flow in lattice QCD,''
  JHEP {\bf 1008} (2010) 071
  [arXiv:1006.4518 [hep-lat]].

\bibitem{Hands:1990wc}
  S.~J.~Hands and M.~Teper,
  ``On The Value And Origin Of The Chiral Condensate In Quenched SU(2) Lattice
  Gauge Theory,''
  Nucl.\ Phys.\  B {\bf 347} (1990) 819.
\bibitem{DeGrand:2000gq}
  T.~A.~DeGrand and A.~Hasenfratz,
  ``Low-lying fermion modes, topology and light hadrons in quenched QCD,''
  Phys.\ Rev.\  D {\bf 64} (2001) 034512
  [arXiv:hep-lat/0012021].
\bibitem{Edwards:1998gk}
  R.~G.~Edwards, U.~M.~Heller and R.~Narayanan,
  ``The hermitian Wilson-Dirac operator in smooth SU(2) instanton
  backgrounds,''
  Nucl.\ Phys.\  B {\bf 522} (1998) 285
  [arXiv:hep-lat/9801015].

\bibitem{Hasenfratz:1998ri}
  P.~Hasenfratz, V.~Laliena and F.~Niedermayer,
  ``The index theorem in QCD with a finite cut-off,''
  Phys.\ Lett.\  B {\bf 427} (1998) 125
  [arXiv:hep-lat/9801021].
\bibitem{Niedermayer:1998bi}
  F.~Niedermayer,
  ``Exact chiral symmetry, topological charge and related topics,''
  Nucl.\ Phys.\ Proc.\ Suppl.\  {\bf 73} (1999) 105
  [arXiv:hep-lat/9810026].
\bibitem{Horvath:2002yn}
  I.~Horvath, S.~J.~Dong, T.~Draper, F.~X.~Lee, K.~F.~Liu, H.~B.~Thacker and J.~B.~Zhang,
  ``On the local structure of topological charge fluctuations in QCD,''
  Phys.\ Rev.\  D {\bf 67} (2003) 011501
  [arXiv:hep-lat/0203027].
\bibitem{Gattringer:2002gn}
  C.~Gattringer,
  ``Testing the self-duality of topological lumps in SU(3) lattice gauge
  theory,''
  Phys.\ Rev.\ Lett.\  {\bf 88} (2002) 221601
  [arXiv:hep-lat/0202002].
\bibitem{Moran:2010rn}
  P.~J.~Moran, D.~B.~Leinweber and J.~Zhang,
  ``Wilson mass dependence of the overlap topological charge density,''
  Phys.\ Lett.\  B {\bf 695} (2011) 337
  [arXiv:1007.0854 [hep-lat]].

\bibitem{Bruckmann:2005hy}
  F.~Bruckmann and E.~M.~Ilgenfritz,
  ``Laplacian modes probing gauge fields,''
  Phys.\ Rev.\  D {\bf 72} (2005) 114502
  [arXiv:hep-lat/0509020].

\bibitem{Solbrig:2007nr}
  S.~Solbrig, F.~Bruckmann, C.~Gattringer, E.~M.~Ilgenfritz, M.~Muller-Preussker and A.~Schafer,
  ``Smearing and filtering methods in lattice QCD - a quantitative
  comparison,''
  PoS {\bf LAT2007} (2007) 334
  [arXiv:0710.0480 [hep-lat]].
\bibitem{Bruckmann:2007ww}
  F.~Bruckmann,
  ``Exploring the QCD vacuum - (some) recent developments in confinement and
  topology,''
  PoS {\bf LAT2007} (2007) 006
  [arXiv:0710.2788 [hep-lat]].
\bibitem{Bruckmann:2009vb}
  F.~Bruckmann, F.~Gruber, C.~B.~Lang, M.~Limmer, T.~Maurer, A.~Schafer and S.~Solbrig,
  ``Comparison of filtering methods in SU(3) lattice gauge theory,''
  PoS  {\bf CONFINEMENT8} (2008) 045
  [arXiv:0901.2286 [hep-lat]].

\bibitem{Gonzalez-Arroyo2006}
A.~Gonz\'alez-Arroyo and R.~Kirchner,
  ``Adjoint modes as probes of gauge field structure,''
  JHEP {\bf 0601} (2006) 029
  [arXiv:hep-lat/0507036].

\bibitem{Novikov:1985ic}
  V.~A.~Novikov, M.~A.~Shifman, A.~I.~Vainshtein, V.~I.~Zakharov,
  ``Supersymmetric Instanton Calculus (Gauge Theories with Matter),''
  Nucl.\ Phys.\  {\bf B260}, 157-181 (1985).

\bibitem{Neuberger:1997fp}
  H.~Neuberger,
  ``Exactly massless quarks on the lattice,''
  Phys.\ Lett.\  B {\bf 417} (1998) 141
  [arXiv:hep-lat/9707022].

\bibitem{latproc}
M.~Garc\'{\i}a Perez, A.~Gonz\'alez-Arroyo and A.~Sastre,
  ``Adjoint zero-modes as a tool to understand the Yang-Mills vacuum,''
  PoS {\bf LAT2007} (2007) 328
  [arXiv:0710.0455 [hep-lat]];

\bibitem{Perez:2010ja}
M.~Garc\'{\i}a Perez, A.~Gonz\'alez-Arroyo and A.~Sastre,
``Probing the Yang-Mills vacuum with adjoint zero-modes,''
  PoS {\bf LATTICE2010} (2010) 285
  [arXiv:1010.6210 [hep-lat]].


\bibitem{alfonsothesis}
Alfonso Sastre. \textit{PhD Thesis: Application of the Dirac operator in the adjoint representation to Yang-Mills theories.}
September 2010, Universidad Aut\'onoma de Madrid


\bibitem{critslow}
S.~Schaefer, R.~Sommer and F.~Virotta  [ALPHA Collaboration],
  ``Critical slowing down and error analysis in lattice QCD simulations,''
  Nucl.\ Phys.\  B {\bf 845} (2011) 93
  [arXiv:1009.5228 [hep-lat]].
\bibitem{critslow2}
 L.~Del Debbio, H.~Panagopoulos and E.~Vicari,
  ``Theta dependence of SU(N) gauge theories,''
  JHEP {\bf 0208} (2002) 044
  [arXiv:hep-th/0204125].
\bibitem{critslow3}
B.~Alles, G.~Boyd, M.~D'Elia, A.~Di Giacomo and E.~Vicari,
  ``Hybrid Monte Carlo and topological modes of full QCD,''
  Phys.\ Lett.\  B {\bf 389} (1996) 107
  [arXiv:hep-lat/9607049].

\bibitem{Neuberger:1998my}
  H.~Neuberger,
  ``A practical implementation of the overlap-Dirac operator,''
  Phys.\ Rev.\ Lett.\  {\bf 81} (1998) 4060
  [arXiv:hep-lat/9806025].


\bibitem{Edwards:1998yw}
  R.~G.~Edwards, U.~M.~Heller and R.~Narayanan,
  ``A study of practical implementations of the overlap-Dirac operator in  four
  dimensions,''
  Nucl.\ Phys.\  B {\bf 540}, 457 (1999)
  [arXiv:hep-lat/9807017].

\bibitem{Jegerlehner:1996pm}
  B.~Jegerlehner,
  ``Krylov space solvers for shifted linear systems,''
  arXiv:hep-lat/9612014.

\bibitem{Neuberger:1998jk}
  H.~Neuberger,
  ``Minimizing storage in implementations of the overlap lattice-Dirac
  operator,''
  Int.\ J.\ Mod.\ Phys.\  C {\bf 10}, 1051 (1999)
  [arXiv:hep-lat/9811019].

\bibitem{Kalkreuter:1995mm}
  T.~Kalkreuter and H.~Simma,
  ``An Accelerated Conjugate Gradient Algorithm To Compute Low Lying
  Eigenvalues: A Study For The Dirac Operator In SU(2) Lattice QCD,''
  Comput.\ Phys.\ Commun.\  {\bf 93} (1996) 33
  [arXiv:hep-lat/9507023].

\bibitem{Borici:1998mr}
  A.~Borici,
  ``On the Neuberger overlap operator,''
  Phys.\ Lett.\  B {\bf 453}, 46 (1999)
  [arXiv:hep-lat/9810064].

\bibitem{kvanbaal1}
 T.~C.~Kraan and P.~van Baal,
 ``Exact T-duality between calorons and Taub - NUT spaces,''
 Phys.\ Lett.\ B {\bf 428} (1998) 268
 [arXiv:hep-th/9802049];
``Periodic instantons with non-trivial holonomy,''
  Nucl.\ Phys.\ B {\bf 533} (1998) 627
  [arXiv:hep-th/9805168];
  ``Monopole constituents inside SU(n) calorons,''
  Phys.\ Lett.\ B {\bf 435} (1998) 389
  [arXiv:hep-th/9806034].

\bibitem{lee1}
K.~M.~Lee,
  ``Instantons and magnetic monopoles on R**3 x S(1) with arbitrary simple
  gauge groups,''
  Phys.\ Lett.\ B {\bf 426} (1998) 323
  [arXiv:hep-th/9802012];
K.~M.~Lee and C.~h.~Lu,
  ``SU(2) calorons and magnetic monopoles,''
  Phys.\ Rev.\ D {\bf 58}, 025011 (1998)
  [arXiv:hep-th/9802108].

\bibitem{'tHooft:1981sz}
  G.~'t Hooft,
  ``Some Twisted Selfdual Solutions For The Yang-Mills Equations On A
  Hypertorus,''
  Commun.\ Math.\ Phys.\  {\bf 81} (1981) 267.

\bibitem{GarciaPerez:1989gt}
 M.~Garc\'{\i}a Perez, A.~Gonz\'alez-Arroyo and B.~Soderberg,
  ``Minimum action solutions for SU(2) gauge theory on the torus with 
  nonorthogonal twist,''
  Phys.\ Lett.\  B {\bf 235} (1990) 117;
M.~Garc\'{\i}a P\'erez and A.~Gonz\'alez-Arroyo,
  ``Numerical study of Yang-Mills classical solutions on the twisted torus,''
  J.\ Phys.\ A  {\bf 26} (1993) 2667
  [arXiv:hep-lat/9206016].



\bibitem{'tHooft:1979uj}
  G.~'t Hooft,
  ``A Property Of Electric And Magnetic Flux In Nonabelian Gauge Theories,''
  Nucl.\ Phys.\  B {\bf 153} (1979) 141.

\bibitem{GonzalezArroyo:1981vw}
  A.~Gonz\'alez-Arroyo, J.~Jurkiewicz and C.~P.~Korthals-Altes,
  ``Ground State Metamorphosis For Yang-Mills Fields On A Finite Periodic
  Lattice,'' Published in Freiburg ASI 1981:0339 (QC175:N15).

\bibitem{Coste:1985mn}
  A.~Coste, A.~Gonzalez-Arroyo, J.~Jurkiewicz and C.~P.~Korthals Altes,
  Nucl.\ Phys.\  B {\bf 262} (1985) 67.


\bibitem{Braam1989}
P.~J.~Braam and P.~van Baal,
  ``Nahm's transformation for instantons,''
  Commun.\ Math.\ Phys.\  {\bf 122} (1989) 267.


\bibitem{Diakonov}
  D.~Diakonov and V.~Petrov,
  ``Confining ensemble of dyons,''
  Phys.\ Rev.\  D {\bf 76} (2007) 056001
  [arXiv:0704.3181 [hep-th]];
  D.~Diakonov, N.~Gromov, V.~Petrov and S.~Slizovskiy,
  ``Quantum weights of dyons and of instantons with non-trivial holonomy,''
  Phys.\ Rev.\  D {\bf 70} (2004) 036003
  [arXiv:hep-th/0404042];
  D.~Diakonov,
  ``Statistical physics of dyons and confinement,''
  Acta Phys.\ Polon.\  B {\bf 39} (2008) 3365
  [arXiv:0807.0902 [hep-th]].

\bibitem{Ilgenfritz}
P.~Gerhold, E.~M.~Ilgenfritz and M.~Muller-Preussker,
  ``An SU(2) KvBLL caloron gas model and confinement,''
  Nucl.\ Phys.\  B {\bf 760} (2007) 1
  [arXiv:hep-ph/0607315].
  a
\bibitem{Bruckmann:2009nw}
  F.~Bruckmann, S.~Dinter, E.~M.~Ilgenfritz, M.~Muller-Preussker and M.~Wagner,
  ``Cautionary remarks on the moduli space metric for multi-dyon simulations,''
  Phys.\ Rev.\  D {\bf 79} (2009) 116007
  [arXiv:0903.3075 [hep-ph]].


\end{thebibliography}
\end{document}